\documentclass[11pt,showpacs,showkeys]{revtex4}
\input epsf.sty

\usepackage{enumerate}
\usepackage{ulem}
\usepackage{cancel}
\usepackage{color}

\def\red#1{\textcolor{red}{#1}}

\def\comment#1{}

\usepackage{graphicx}
\begin{document}

\title{Cosmological $\Lambda$ driven inflation and produced massive particles}
\author{She-Sheng Xue}
\email{xue@icra.it; shesheng.xue@gmail.com}
\affiliation{ICRANet, Piazzale della Repubblica, 10-65122, Pescara,\\
Physics Department, Sapienza University of Rome, P.le A. Moro 5, 00185, Rome,
Italy} 


\begin{abstract}
Suppose that the early Universe starts with a quantum spacetime originated cosmological $\Lambda$-term at the Planck scale $M_{\rm pl}$. The cosmological energy density $\rho_{_{_\Lambda}}$ drives inflation and simultaneously
reduces its value to create the matter-energy density $\rho_{_{_M}}$ via the continuous pair productions of massive fermions and antifermions. 
The decreasing $\rho_{_{_\Lambda}}$ and increasing $\rho_{_{_M}}$, in turn, slows down the inflation to its end when the pair production rate $\Gamma_M$ is larger than the Hubble rate $H$ of inflation. 
Such back-reaction evolutions of the density $\rho_{_{_\Lambda}}$ 
and Hubble rate $H$ are uniquely determined by two independent equations from the Einstein equation and energy conservation law, besides, 
the $\rho_{_{_M}}$ and its equation of state as functions of $H$ are determined by continuous massive pair productions. For very massive and dense 
pairs $m\gg H$, $\rho_{_{_M}}\propto m^2H^2$ and $\rho_{_{_\Lambda}}\propto M^2_{\rm pl}H^2>\rho_{_{_M}}$. As a result, inflation naturally appears and theoretical results agree with Planck 2018 observations. The CMB large-scale anomaly can be possibly explained and the dark-matter acoustic wave is speculated. Suppose that the reheating efficiently converts the cosmological energy density $\rho_{_{_\Lambda}}$ to the matter-energy density $\rho_{_{_M}}\gg \rho_{_{_\Lambda}}$ accounting for the most relevant Universe mass, and some massive pairs decay to relativistic particles of energy density $\rho_{_{_R}}$ starting the hot Big Bang. Since then, the energy density $\rho_{_{_R}}$ produced at the reheating predominately governs the decreasing Hubble rate $H^2\propto \rho_{_{_R}}$, and massive pair productions are small and unimportant. However, the aforementioned 
back reaction $\rho_{_{_M}}\leftrightarrow H\leftrightarrow \rho_{_{_\Lambda}}$
is weak but continues in standard cosmological evolution. As a consequence, the cosmological energy density $\rho_{_\Lambda}$ closely tracks down the energy density $\rho_{_R}$ from the reheating end up to the radiation-matter equilibrium, then it varies very slowly, $\rho_{_\Lambda}\propto$ constant, due to the transition from radiation dominate to matter dominate epoch. Therefore the cosmic coincidence problem can be possibly avoided.   
The relation between $\rho_{_\Lambda}$ and radiation and matter-energy densities is obtained and can be examined at large redshifts.  
\end{abstract}

\maketitle

\newpage
\tableofcontents
\newpage

\section{\bf Introduction}\label{introduction}
In the standard model of modern cosmology ($\Lambda$CDM), the cosmological constant, inflation, reheating, dark matter and coincidence problem have been long-standing basic issues for decades. The inflation \cite{inflation0} is a fundamental epoch and the reheating \cite{RevReheating} is a critical mechanism, which transitions the Universe from the cold massive state left by inflation to the hot Big Bang \cite{BigBang}. The cosmic microwave background (CMB) observations have been attempting to determine a unique model of inflation and reheating.
On the other hand, what is the crucial role that 
the cosmological $\Lambda$ term play in inflation and reheating, and what is the essential reason for the coincidence of dark-matter dominate matter density and the cosmological $\Lambda$ energy density. 
There are various models and many efforts, 
that have been made to approach these issues, and readers are referred to 
review articles and professional books, for example, 
see Refs.~\cite{Peebles,kolb,book,Inflation_R,
Inflation_higgs,reviewL,Bamba:2012cp,Nojiri:2017ncd,
Coley2019,
prigogine1989,prigogine1989+,wangbin,
wuyueliang2012,
wuyueliang2016,axioninf,xueNPB2015}. 

We attempt in this article to give an insight into some points of these issues. 
Suppose that the quantum gravity originates the cosmological 
term $\Lambda\sim M^2_{\rm pl}$ at the Planck scale. 
The initial state of the Universe is an approximate de Sitter spacetime of the horizon 
$H_\circ\approx (\Lambda/3)^{1/2}$ without any matter. 
The cosmological $\Lambda$ energy 
density drives the spacetime inflation with the scale factor 
$a(t)\approx e^{H_\circ t}$. On the other hand, de Sitter spacetime is unstable 
against spontaneous particle creations \cite{pairproduction, pairproductionCold}. By reducing its value, the cosmological $\Lambda$ term creates 
very massive pairs of fermions and antifermions $m \sim M_{\rm pl}$ 
for matter content. We adopt three equations of Einstein, conservation law and 
produced pair density to determine 
the cosmological energy density $\rho_{_{_\Lambda}}$ governed the spacetime 
inflation rate $H$ and in the meantime produced the matter density $\rho_{_{_M}}$, 
whose back reaction, in turn, slows down inflation, comparing with 
CMB observations. 
Analogously, suppose that after reheating the matter-energy density is much larger than 
the cosmological energy density, we examine whether such back reaction 
links two densities in Universe expansion, consistently leading to the cosmic
coincidence in the present time, briefly presented in Ref.~\cite{xue2020MPLA}.

We organise this article as follow. we revisit in Sec.~\ref{Ein} the 
Einstein equation and 
conservation law in the view of 
time-varying 
cosmological $\Lambda$-term 
coupling with the matter. 
In Sec.~\ref{production}, 
we present the discussions and calculations of the matter produced from 
the spacetime through the 
pair-production of fermions and antifermion.
Based on these results and equations, 
we adopt numerical and analytical approaches to study the inflation epoch in connection with observations in Sec.~\ref{inflation}. 
We study the relationships of the horizon $H$, the cosmological 
$\Lambda$-term and matter term varying in time after Big Bang, particularly focusing on the problem of cosmic coincidence in Sec~\ref{coincidence}. 
A summary and some remarks are given in the concluding section. 
   
\section{\bf Einstein equation and generalized conservation law}\label{Ein}

\subsection{The role of cosmological $\Lambda$-term}

The Einstein equation for 
the spacetime of Einstein tensor 
${\mathcal G}^{ab}$ coupling to the matter of energy-momentum 
tensor $T^{ab}_{_M}$ reads, see for example Ref.~\cite{Weinberg1972},
\begin{eqnarray}
{\mathcal G}^{ab} = -8\pi 
G T^{ab}_{_M}; \quad
{\mathcal G}^{ab} =  R^{ab} -(1/2) g^{ab}R -\Lambda g^{ab},
\label{e1}
\end{eqnarray}
where $R^{ab}$ ($R$) is the Ricci tensor (scalar), and $G$ is the Newton constant.
Its covariant differentiation and the Bianchi identity are 
\begin{eqnarray}
{\mathcal G}^{ab}_{\,\,\,\,\,\,;\,b} = -8\pi 
\,G \,T_{_M}^{ab}\,_{\,;\,b},
\quad
[ R^{ab}-(1/2)\delta^{ab}R]_{\,;\,b}\equiv 0,
\label{de1}
\end{eqnarray}
which lead to the generalized conservation law \cite{xueNPB2015},
\begin{equation}
(\Lambda)_{;\,b}\,g^{ab}=
8\pi G (T^{ab}_{_M})_{;\,b}\,\, ,
\label{geqi00}
\end{equation} 
with time-varying cosmological term $(\Lambda)_{;\,b}=(\Lambda)_{,\,b}$. 
Equation (\ref{geqi00}) clearly shows that the cosmological 
$\Lambda$-term explicitly couples with the matter $T^{ab}_{_M}$ of produced pairs. 

Assume that produced pairs are so dense and massive, 
their density $\rho_{_{_M}}$ and pressure $p_{_{_M}}$ 
represent semi-classical averages over many pairs production, 
and their energy-momentum tensor is approximately described as a perfect fluid, 
\begin{eqnarray}
T^{ab}_{_M} &=& p_{_{_M}} g^{ab} +(p_{_{_M}}  + \rho_{_{_M}}) U^aU^b,
\label{emtm}
\end{eqnarray} 
with respect to the observer of the four velocity $U^a=(1, 0,0,0)$.  
\comment{
As will be shown, the matter-energy density $\rho_{_{_M}}=\rho_{_{_M}}(H)$ and pressure $p_{_{_M}}=p_{_{_M}}(H)$ as functions of the horizon $H$ are uniquely 
calculated by using the Hawking-Parker type pair production processes. 
The most relevant amount of the matter in the Universe is 
created in the inflation \red{and reheating} epochs of early Universe \red{by the conversion from the cosmological energy density $\rho_{_{_\Lambda}}$}. As a consequence, 
the time evolutions of the horizon $H$ 
and the cosmological constant $\Lambda$ are completely determined 
by the two Einstein equations, provided their initial values are given. 
The obtained analytical solutions and numerical results are compared with observations.
}

Moreover, as will be shown in Sec.~\ref{production}, 
the cosmological $\Lambda$-term implicitly couples with the matter $T^{ab}_{_M}(H)$ through the production and annihilation of 
particle-antiparticle pairs via the horizon $H$ of the spacetime, 
which is in turn 
governed by the Einstein equation (\ref{e1}).

Despite its {\it essence} of spacetime origin, 
the cosmological $\Lambda$-term 
in the Einstein spacetime tensor ${\mathcal G}_{ab}$ 
can be moved to the RHS of Einstein equation (\ref{e1}), 
and {\it formally} expressed by 
using a {\it symbol} of energy-momentum tensor 
$T^{ab}_{_{_{\Lambda}}}$, analogously to the matter $T^{ab}_{_{M}}$ (\ref{emtm}),
\begin{eqnarray}
T^{ab}_{_{\Lambda}} &\equiv &p_{_{_{\Lambda}}} g^{ab} +(p_{_{_{\Lambda}}}  + \rho_{_{_{\Lambda}}} ) U^aU^b\equiv -\rho_{_{_{\Lambda}}} g^{ab},\label{emt}
\end{eqnarray} 
and implementing a negative mass density 
$\rho_{_\Lambda}\equiv \Lambda/(8\pi 
G)\equiv -p_{_\Lambda}$ identically. This practical 
analogy between 
$T^{ab}_{_{\Lambda}}$ (\ref{emt}) and $T^{ab}_{_{\Lambda}}$ (\ref{emtm}) is purely 
{\it technical} in the sense of the convenience for calculations and expressions 
below. In so doing, we do not make any model to change the spacetime nature of the cosmological $\Lambda$-term. The interested readers are referred to the 
Ref.~\cite{xuecos2009} for the more detailed discussions on the cosmological $\Lambda$-term with respect to the vacuum energy of local field theories.   

Using the technical notation $T^{ab}_{_{\Lambda}}\equiv -\rho_{_\Lambda}g^{ab}$ 
(\ref{emt}), from Eqs.~(\ref{e1}) and (\ref{de1}) it can be derived that the 
generalized conservation law (\ref{geqi00}) is equivalent to the 
conservation law expressed by 
\begin{eqnarray}
(T^{ab})_{\,\,;\,b}= 0;
\quad T^{ab}\equiv T^{ab}_{_M} + T^{ab}_{_\Lambda},
\label{totalc}
\end{eqnarray}
in terms of $T^{ab}_{_M}$ and $T^{ab}_{_\Lambda}$. The generalized conservation 
law (\ref{geqi00}) or (\ref{totalc}) represents the coupling relationship 
among the cosmological 
$\Lambda$-term and matter M-term,  
all of them are varying in time. Equation (\ref{geqi00}) or (\ref{totalc}) 
is one of fundamental equations studied in the present article. 
Note that the generalized conservation law Eq.~(\ref{geqi00}) or 
(\ref{totalc}) reduces to the usual matter conservation law 
$(T^{ab}_{_M})_{\,\,;\,b}=0$ in the the $\Lambda$CDM model 
of the constant cosmological $\Lambda$-term $(\Lambda)_{,\,b}=0$.

\subsection{Generalized equations for Friedmann Universe}\label{gFriedmann}

In this section, following the general equations  (\ref{e1}), (\ref{de1}) 
and (\ref{geqi00}) previously discussed, 
we derive the generalized equations describing the 
evolution of Friedmann Universe. These are basic equations in this article 
that we use to study the inflation, reheating, radiation and matter 
dominated epochs in Universe evolution.  

\subsubsection{Generalized equations}

In the Robertson-Walker spacetime $ds^2=-dt^2+a^2(t)d{\bf x}^2$ of zero spatial curvature and scaling factor $a(t)$, 
Equations (\ref{e1},\ref{de1}) and (\ref{geqi00}) 
become the following 
equations \cite{xueNPB2015}. 

The first equation comes from the 
$0\!-\!0$ component of Einstein equation (\ref{e1}),
\begin{eqnarray}
h^2 = 
(\Omega_{_M} + \Omega_{_\Lambda}),~~ h\equiv  H/H_{\rm ch}, 
~~ \Omega_{_{M,\Lambda}}\equiv \rho_{_{M,\Lambda}}/\rho_c^{\rm ch}, ~~~
\label{e3}
\end{eqnarray}
where $H_{\rm ch}$, $a_{\rm ch}$ and $\rho_c^{\rm ch}\equiv 3H_{\rm ch}^2/(8\pi G)$ represent 
the characteristic horizon scale, scaling factor and critical density 
of Universe evolution at each epoch, i.e., inflation, reheating, radiation 
or matter dominated epoch. They should eventually be determined by observations.  
As an example, in the present Universe, $H_{\rm ch}=H_0$, $a_{\rm ch}=a_0$,
$\rho_c^{\rm ch}=\rho_c\equiv 3H_0^2/(8\pi G)$ and $G$ is the value 
of gravitational Newton constant today. 

The second equation comes from the 
$1\!\!-\!\!1$ component of Einstein equation (\ref{e1})
\begin{eqnarray}
\frac{dh^2}{dx}\! +\! 2h^2 \!=\! \frac{2\ddot a}{H_{\rm ch}^2a}\!=\!
\Big[2\Omega_{_\Lambda}\!-\!(1\!+\!3\omega_{_M})\Omega_{_M}\Big],
\label{e2}
\end{eqnarray}
where $\omega_{_M} = p_{_M}/\rho_{_M}$ is the equation of state for the matter content, and we introduce the variable $\quad x=\ln(a/a_{\rm ch})$ and derivative 
$d(\cdot\cdot\cdot)/dt=Hd(\cdot\cdot\cdot)/dx$ for convenience later on.
The first and second equations (\ref{e3}) and (\ref{e2}) reduce to 
the corresponding Friedmann equations.

The third equation is derived from the generalized conservation 
law (\ref{geqi00}) or (\ref{totalc}) in virtue 
of Eqs. ~(\ref{emtm}) and (\ref{emt}),  
\begin{eqnarray}
\frac{d\Lambda}{dt}
&=&-8\pi G\left[ \frac{d\rho_{_M}}{dt}+ \frac{3\dot a}{a}(p_{_M}+\rho_{_M})\right],
\label{cgeqi200}
\end{eqnarray}
which can be recast as
\begin{eqnarray}
\frac{d}{dx}\Big[(\Omega_{_\Lambda}+\Omega_{_M})\big]=-3(1+\omega_{_M})\Omega_{_M}.
\label{cgeqi20}
\end{eqnarray} 
It reduces to the normal conservation law for the matter 
\begin{eqnarray}
\frac{d\rho_{_M}}{dt}+\frac{3\dot a}{a}(p_{_M}+\rho_{_M})=0,~{\rm i.e.,}~\frac{d}{dx}\Omega_{_M}=-3(1+\omega_{_M})\Omega_{_M},
\label{cgeqi2m}
\end{eqnarray}
when the cosmological term $\Lambda$ 
is constant in time. 
The combination of Eqs.~(\ref{e3}) and (\ref{cgeqi20}) yields Eq.~(\ref{e2}), 
similarly to the case of the usual Friedmann equations. 
\comment{The fundamental and independent equations used in this article 
are Eqs.~(\ref{e3}) and (\ref{cgeqi20}) only, reducing to the standard 
Friedmann equations for $\Omega_{_\Lambda}={\rm constant}$.  
We do not use Eq.~(\ref{e2}), since it is not an independent equation. However, Equation (\ref{e2}) shows that as dynamically varying $\Omega_{_\Lambda}$ and $\Omega_{_M}$, the Universe accelerates $\ddot a\propto 2\Omega_{_\Lambda}\!-\!(1\!+\!3\omega_{_M})\Omega_{_M}>0$, decelerates $\ddot a\propto 2\Omega_{_\Lambda}\!-\!(1\!+\!3\omega_{_M})\Omega_{_M}<0$ and turns from one to another at $\ddot a\propto 2\Omega_{_\Lambda}\!-\!(1\!+\!3\omega_{_M})\Omega_{_M}=0$. 
}

\comment{
In order to use two ind~ependent equations (\ref{e3}) 
and (\ref{cgeqi20}) to determine the time evolutions of 
the Hubble horizon $H$ and cosmological $\Lambda$-term in 
Universe evolution, we have to know at least 
two additional equations or formula to describe: 
\begin{enumerate}[(i)] 
\item how the matter $\Omega_{_M}$ is produced and/or evolves in time, 
as a function of $H$ or $\Omega_{_\Lambda}$.
This will be discussed in Sec.~\ref{production};
\item  how the gravitational coupling $g$ varies in time, as a function of the scaling factor $a$. 
This can be obtained from the renormalization group equation at an adequate fixed point, see for example Refs.~\cite{xueNPB2015} and \cite{JS2016}. \label{g=1}
\end{enumerate}
}
It is worthwhile to stress again that in Eqs.~(\ref{e3}) 
and (\ref{cgeqi20}) we treat the Hubble horizon $H$ and 
cosmological $\Lambda$-term as primarily physical quantities describing 
the spacetime nature, whose quantum origin at the Planck scale 
is not the topic of this article.     


\subsubsection{Two independent equations for uniquely determining $H$ and $\Lambda$}


In summary, Einstein equations (\ref{e3},\ref{e2}) and (\ref{cgeqi20}) are recast as 
the following set of two independent equations,
\begin{eqnarray}
&& h^2 = (\Omega_{_M} + \Omega_{_\Lambda}),
\label{fe3}\\
&&\frac{d}{dx}\left(\Omega_{_\Lambda}+\Omega_{_M}\right)
=-3(1+\omega_{_M})\Omega_{_M},
\label{fcgeqi20}
\end{eqnarray}
where the first equation is the Friedmann equation and the second 
equation is the energy conservation law of spacetime and matter, 
generalized from the usual matter conservation law (\ref{cgeqi2m}).

However, the numbers of Eqs.~(\ref{fe3}) and (\ref{fcgeqi20}) are not enough to completely 
determine unknown quantities of the Hubble horizon $h$, 
cosmological term $\Omega_{_\Lambda}$ and the matter term $\Omega_{_M}$, as well as $\omega_{_M}$
the equation of state of the matter. 
Our approach to this issue is that the matter has been produced via the 
process from the 
space time of the horizon $H$, since the beginning of the Universe, 
the matter energy density $\rho_{_{_M}}=\rho_{_{_M}}(H)$ 
and pressure $p_{_{_M}}=p_{_{_M}}(H)$ are uniquely calculated 
as functions of the horizon $H$,  
\begin{eqnarray}
\Omega_{_M}=\Omega_{_M}(H);\quad \omega_{_M}\equiv p_{_{_M}}/\rho_{_{_M}}=\omega_{_M}(H)
\label{omeh}
\end{eqnarray}
as shown in the next Sec.~\ref{production}. 
As a result, Einstein equations (\ref{fe3}) and
(\ref{fcgeqi20}) together with the relationship (\ref{omeh}) from the 
pair production process  
lead to a set of complicated nonlinear back-reacting equations, which however
completely determine the $H$, $\Omega_{_\Lambda}$ and $\Omega_{_M}$ 
governing the evolution of the Universe, 
provided that their initial conditions are given in each epoch.

\comment{ In some literatures,  
a new arbitrary function $q(x)$ is introduced to describe the $\Omega_{_\Lambda}-\Omega_{_M}$ interaction and split Eq.~(\ref{fcgeqi20}) 
into two equations
\begin{eqnarray}
\frac{d\Omega_{_M}}{dx}
=-3(1+\omega_{_M})\Omega_{_M} + q(x);\quad
\frac{d\Omega_{_\Lambda}}{dx}=-q(x),
\label{twoq}
\end{eqnarray}
}

\subsubsection{Initial condition and scale}

We are going to use differential equations (\ref{fe3}) and (\ref{fcgeqi20}) to study  
each epoch of the Universe evolution: the inflation, reheating, radiation and matter dominated epoch. 
For simplicity and convenience in notations, the characteristic 
horizon scale $H_{\rm ch}$ and scaling factor $a_{\rm ch}$ (\ref{e3}) are chosen 
as initial scales in each epoch. Thus, we use the index ``$i$'' to 
indicate each epoch and its initial conditions, 
which can be generally indicated by the scaling factor $a^i_{\rm ch}$, 
\begin{eqnarray}
&&{\rm Hubble~rate}~ H^i_{\rm ch},\quad h^i=H^i/H^i_{\rm ch},\quad {\rm critical~density}~\rho^i_c=3(H^i_{\rm ch})^2/(8\pi M^{-2}_{\rm pl});\label{initH}\\
&&{\rm cosmological}~ \Lambda\!\!-\!\!{\rm term}~\Omega^i_{_\Lambda}=\rho^i_{_\Lambda}/\rho^i_c,\quad {\rm matter~content}~~ \Omega^i_{_M}=\rho^i_{_M}/\rho^i_c;\label{initLM}
\end{eqnarray}
to describe the beginning of each epoch under consideration. These initial scales (\ref{initH}) and 
(\ref{initLM}) relate to not only the characteristic scale (\ref{e3}) of each 
epoch in Universe evolution, but also the transitions from one epoch to another. 
We will duly specify these initial conditions and scales in discussing each 
epoch of Universe evolution. On the other hand, these initial conditions and scales 
should be chosen in the range where has the validity of the effective 
theory (\ref{emtm}) and equations (\ref{e3},\ref{cgeqi20}) 
describing the Universe evolution. For instance, the initial scales for 
the inflation epoch should be smaller than the Planck scale. 

In the following way, we will try to solve 
equations (\ref{fe3}) and (\ref{fcgeqi20}). 
Starting from the initial values 
$\Omega^i_{_\Lambda}$ and $\Omega^i_{_M}$ (\ref{initLM})
at an adequate characteristic scale $H_i$ (\ref{initH}), 
$\Omega_{_\Lambda}(h)$ and $\Omega_{_M}(h)$ govern the varying
spacetime horizon $h$. The variation $\Omega_{_M}(h)$ dynamically
leads to the variation of $h^2$ and $\Omega_{_\Lambda}$ 
via Eq.~(\ref{fcgeqi20}), in turn $\Omega_{_M}(h)$ changes via Eq.~(\ref{fe3}). 
This completely determines $h^2(x)$ and $\Omega_{_\Lambda}(h)$ scaling in the Universe evolution, provided that $\Omega_{_M}(h)$ and $\omega_{_M}(h)$ (\ref{omeh}) are calculated 
as functions of $h$ via the pair production process. 

In this article, it is useful to introduce the $\epsilon$-rate of $H$-variation:
\begin{eqnarray}
\epsilon &\equiv&  - \frac{\dot H}{H^2}= - \frac{1}{H}\frac{d H}{dx}
= - \frac{1}{h}\frac{d h}{dx}
\label{erate0}\\
&=& \frac{3}{2}(1+\omega_{_M})\frac{\Omega_{_M}}{\Omega_{_\Lambda}+\Omega_{_M}}
\label{erate0+}
\end{eqnarray}
to characterize different epochs of the Universe evolution.  
As a convenient unit for calculations and expressions, 
we adopt the reduced Planck scale $m_{\rm pl}\equiv (8\pi G)^{-1/2}=1$,
unless otherwise stated. Note that the reduced Planck scale 
$m_{\rm pl}=(8\pi)^{-1/2} M_{\rm pl}=2.43\times 10^{18} $GeV.



\section{\bf Pair production from spacetime}\label{production}
In this section, we describe how the matter is produced from the spacetime by the pair production of fermions $F$ and antifermions $\bar F$, and discuss how to calculate the matter content $\Omega_{_M}(h)$ as a function of $h$ or $\Omega_{_\Lambda}(h)$. This is crucial to study and solve the 
generalized Friedmann equations (\ref{fe3}) and (\ref{fcgeqi20}).

\subsection{Matter produced from spacetime}
The matter production from the spacetime is attributed to the 
production of fermions $F$ and antifermions $\bar F$ 
\begin{eqnarray}
{\mathcal S} \Rightarrow F + \bar F,
\label{spro}
\end{eqnarray}
where ${\mathcal S}$ stands for the spacetime.
Such pair production is a semi-classical process of producing particles and antiparticles in an external and classical field $H$, i.e., 
the horizon of the spacetime, 
which obeys classical equation (\ref{fe3}) 
and (\ref{fcgeqi20}). 

\subsubsection{Fermion-antifermion pair production density in De Sitter spacetime}

{\it A priori}, we assume that the $H$-field varies more 
slowly, compared with the rates of 
pair-production and/or other microscopic processes, namely the 
$\epsilon$-rate (\ref{erate0}) is very small ($\epsilon \ll 1$). 
Therefore, to approximately calculate the matter content 
$\Omega_{_M}(h)$ in the Einstein equations (\ref{fe3}) and (\ref{fcgeqi20}), we consider the spontaneous pair production of 
massive spin-$1/2$ particles $F$ and antiparticles $\bar F$ 
from the exact De Sitter spacetime ${\mathcal S}$
of the constant horizon $H$ and scaling factor $a(t)=e^{Ht}$.

On the basis of semi-classical calculations at the zeroth 
adiabatic order, the averaged number density 
of all pairs produced from the initial time 
$t_i=0$ to the final time $t\gtrsim 2\pi H^{-1}$ is given by Refs.~\cite{lnt2014,ekhard},
\begin{eqnarray}
n_{_M} &=& \frac{H^3}{2\pi^2} \int_0^\infty dz z^2 |\beta^{(0)}_k(t)|^2\nonumber\\
&=&\frac{H^3e^{\pi\mu}}{16\pi} \int_0^\infty dz \frac{z^3}
{\sqrt{z^2+\mu^2}} {\mathcal F}^{(0)}_\nu(z,\mu),
\label{nden}
\end{eqnarray}
and 
\begin{eqnarray}
{\mathcal F}^{(0)}_\nu(z,\mu) &=&\Big|\sigma_+H^{(1)}_{\nu-1}(z)-i\sigma_-H^{(1)}_{\nu}(z)\Big|^2 ,~~~~~
\label{fden}
\end{eqnarray}
$\sigma_\pm\equiv [(z^2+\mu^2)^{1/2}\pm\mu]^{1/2}$, $\nu=1/2-i\mu$, and $H^{(1)}_{\nu}(z)$ is the Hankel function of the first kind.
In Eqs.~(\ref{nden}) and (\ref{fden}), $\mu=m/H$ and $m$ stands for effective particle mass,
the variable $z\equiv kH^{-1}e^{-Ht}=k_{\rm phy}/H$ and 
$k_{\rm phy}=k/a(t)$, relating to the produced particle 
comoving (physical) momentum $k$ ($k_{\rm phy}$). 
The pair-production probability is given by 
the Bogolubov coefficient squared $|\beta^{(0)}_k(t)|^2$ 
and $|\beta^{(0)}_k(t)|^2\sim {\mathcal O}(1/k^{4})$ for large $k$. 
This physically shows that the pair productions from large momentum $k$ 
modes are suppressed, and the semi-classical pair density (\ref{nden}) 
is convergent. 

The semi-classical result (\ref{nden}) is valid only for 
massive particles $m > H$ and their wavelengths $m^{-1} < H^{-1}$, 
namely particles produced are well inside the Horizon. The validity of semi-classical calculations cannot be trusted for light particles ($m<H$), 
whose wavelengths are larger than the horizon size. 

In the exact De Sitter spacetime, $H$ and 
$\Lambda$ are strictly constants in time. 
The number density $n_{_M}$ (\ref{nden}) of produced $F\bar F$ pairs as a function of $H/m$ does not 
depend on the time. This physically implies that the increasing number of pairs produced is compensated by the effects of the expansion of the spacetime.  

\subsubsection{Pairs' energy density and pressure in De Sitter spacetime}

\comment{
Whereas   
the energy-momentum tensor of these pairs is given by
$T^{ab}_{_{D\!S}}=\rho_{_{D\!S}} g^{ab}$ and negative 
energy density $\rho_{_{D\!S}}<0$. 
One may regard $T^{ab}_{_{D\!S}}$ as the energy-momentum tensor 
of a perfect fluid of these pairs with equation of 
state $\rho_{_{D\!S}}+p_{_{D\!S}}=0$ in a macroscopic and 
thermodynamical description. This can be explained as follow.
}

The vacuum Einstein equation ${\mathcal G}^{ab}=0$,
see Eq.~(\ref{e1}), possesses the De Sitter symmetry, i.e., 
has the maximally symmetric solution 
$R^{ab}=-\Lambda g^{ab}$, $R=-4\Lambda$, 
and $\Lambda \equiv 3 H^2$. 
The energy-momentum tensor of these pairs must take 
the form $T^{ab}_{_{DS}}=\rho_{_{DS}} g^{ab}$, 
due to the exact De Sitter symmetry preserved 
in pair productions,
where $\rho_{_{DS}}$ is the pair energy density in De Sitter spacetime. 
If one regards $T^{ab}_{_{DS}}$ as the energy-momentum tensor 
of a perfect fluid, analogously to Eq.~(\ref{emtm}), 
in a macroscopic description of these pairs,
one is led to the equation of 
state $\rho_{_{DS}}+p_{_{DS}}=0$ and negative 
pair energy density $\rho_{_{DS}}<0$.

The negative pair energy density $\rho_{_{DS}}<0$ represents that the 
the pair-production system gains energy at the expense of the spacetime gravitational energy,
in contrast with the normal positive kinetic energy of pairs 
(or thermal energy of pair gas). This means that the spontaneous production 
of pairs is energetically favourable. 
However, the corresponding positive pressure $p_{_{DS}}=-\rho_{_{DS}}>0$ 
to maintain the De Sitter spacetime of exact constant $H$ and $\Lambda$, so that the number density $n_{_M}(\mu)$ (\ref{nden}) does not change in time.

This can be possibly understood by the analogy of the first thermodynamics law for the adiabatic transformation
of the system of the volume $V$
in which the particle number changes in time, see for example Ref.~\cite{prigogine1989},   
\begin{eqnarray}
dQ=d(\rho V) +p dV -[(\rho+p)/n] d(nV)=0,  
\label{tdnt}
\end{eqnarray}
which is equivalent to $d\rho=(\rho+p)dn/n$ or $p=(nd\rho -\rho dn)/dn$ with the particle number density $n$, energy density $\rho$ and normal thermal pressure 
$p$. The third term represents the energy gain by the system due to the change 
in the particle number $nV$. The thermal pressure $p$ is determined by the energy production $d\rho$ and 
particle production $dn$. Equation (\ref{tdnt}) can be rewritten as 
$d(\rho V) +(p+p_n) dV =0$, where 
\begin{eqnarray}
p_n=-\frac{\rho +p}{n} \frac{d(nV)}{dV}.  
\label{tdnt1}
\end{eqnarray}
Suppose that the system is in a thermal equilibrium of the temperature $T$, 
undergoes an adiabatic process $dQ=0$ 
and the number density $n$ of particles 
does not change in time, with further assumptions of  
thermal temperature, pressure and energy density vanishing, i.e., $p=0$ 
and $\rho=\rho_{\rm thermal}+\rho_n=\rho_n$. 
Equation (\ref{tdnt1}) yields $p_n=-\rho_n$, where the negative 
energy density $\rho_n<0$ corresponds to the energy gain in particle productions. 

Moreover, the application of the second 
law of thermodynamics by considering the entropy $S$ 
production due to particle productions leads to
\begin{eqnarray}
TdS&=&d(\rho V) +p dV-\mu d(n V)\nonumber\\
&=&T(S/nV)d(nV)>0,~\Rightarrow ~ S=nV,  
\label{tdnt2}
\end{eqnarray}
where the chemical potential $\mu=\rho+p -TS/V$. This shows that the entropy increases as the number of 
particles produced, implying the processes of    
particle productions are not only energetically, but also entropically favourable.

\subsubsection{Back reaction of pair productions}

However, the cosmological term $\Lambda$ or horizon $H$ must changes/decreases in pair productions, because 
gravitational energy of the spacetime has to pay for the energy gain due to massive pair production and pairs' 
kinetic (thermal) energy. Besides, the microscopic inverse process 
\begin{eqnarray}
F + \bar F \Rightarrow {\mathcal S} 
\label{ispro}
\end{eqnarray}
of particle $F$ and antiparticle $\bar F$ annihilation to the 
spacetime takes place. These back reactions of pair productions on 
the spacetime must be taken into account and 
the cosmological term $\Lambda$ or horizon $H$ are no longer constants but decrease their values
following the Einstein equation (\ref{fe3}) and generalized 
conservation law (\ref{fcgeqi20}), 
in which the matter of 
produced pairs, in turn, acts on the cosmological term $\Lambda$ 
or horizon $H$. In this case, it is no longer an exact 
De Sitter spacetime and the energy-momentum tensor of produced pairs 
is not of the form $T^{ab}_{_{D\!S}}=\rho_{_{D\!S}} g^{ab}$. However, it is not easy 
in the horizon $H$ varying case to exactly 
calculate the energy-momentum tensor of produced pairs, including
 also their thermal energy and pressure. 
In the next section, assuming the very slowly varying horizon $H$, 
we propose an approximate way for calculations and check 
its validity {\it a poteriori}.


\subsection{Fermion-antifermion Pairs' energy-momentum tensor}\label{pairs}

Because of fermion and antifermion pairs $F\bar F$ production (\ref{spro}) and 
annihilation (\ref{ispro}), as well as their backreaction on the spacetime through 
the Eqs.~(\ref{fe3}) and (\ref{fcgeqi20}), the horizon $H$ can not keep exact constancy.
Such back reaction processes lead to a possibly slowly decreasing $H$, 
as a result, the exact De Sitter symmetry of $(H={\rm const})$ is broken. 
As shown in the previous section, such a back-reaction process is energetically 
and entropically favourable. To take into account the back-reaction of
all produced pairs on the Einstein equation, we need to calculate in the 
energy-momentum tensor $T^{ab}_{_M}$ (\ref{emtm}) the energy density 
$\rho_{_M}$ and pressure $p_{_M}$,  
attributed to all produced pairs of the averaged number density $n_{_M}$ (\ref{nden}). 

In this article, we mainly consider the productions of massive pair 
$m \lesssim M_{\rm pl}$ and the slowly decreasing $H$ characterised by the Horizon $H$
and its time variation $\tau^{-1}_{_H}\equiv H^{-1}\dot H$ being much small than the 
pair mass $m> H$ and $m\gg \tau^{-1}_{_H}$,
\begin{eqnarray}
\frac{\tau^{-1}_{_H}}{m}= \frac{H}{m} \epsilon \ll 1,
\label{hrate}
\end{eqnarray} 
where $\epsilon \leq 2$ is the $\epsilon$-rate (\ref{erate0}). Inside 
the Hubble horizon, many pairs have been produced and accumulated from the initial time 
to the Hubble time $t\gtrsim 2\pi H^{-1}$, and pair density is very large. 
In this case,
we calculate in the energy-momentum tensor
$T^{ab}_{_M}$ (\ref{emtm}) the pair energy density $\rho_{_M}$ and pressure $p_{_M}$ 
in the following semi-classical approximation, correspondingly to the averaged pair 
number density (\ref{nden}). The pair energy density and pressure contributed 
from all produced pairs are given by 
\begin{eqnarray}
\rho_{_M} &=& 2\,\frac{H^3}{2\pi^2} \int_0^{z_{\rm cut}} dz z^2
\epsilon_k|\beta^{(0)}_k(t)|^2\nonumber\\
&=& 2\,\frac{H^4e^{\pi\mu}}{16\pi} \int_0^{z_{\rm cut}} dz z^3 {\mathcal F}^{(0)}_\nu(z,\mu) ,
\label{rden}
\end{eqnarray}
and
\begin{eqnarray}
p_{_M} &=& 2\,\frac{H^3}{2\pi^2} \int_0^{z_{\rm cut}} dz z^2 \frac{(k/a)^2}{
3\epsilon_k}|\beta^{(0)}_k(t)|^2\nonumber\\
&=& \frac{\rho_{_M}}{3}
-2\,\frac{\mu^2H^4e^{\pi\mu}}{3\times 16\pi} \int_0^{z_{\rm cut}} dz \frac{z^3}
{z^2+\mu^2} {\mathcal F}^{(0)}_\nu(z,\mu),~~~~~
\label{pden}
\end{eqnarray} 
where the energy spectrum of created particles is
\begin{eqnarray}
\epsilon_k=a^{-1}[(k/a)^2+m^2]^{1/2},
\label{pairspe}
\end{eqnarray}
including both mass and kinetic energy. The equation of state of these pairs is
\begin{eqnarray}
\omega_{_M}=p_{_M}/\rho_{_M}\label{eosf},
\end{eqnarray} 
and the sound velocity $c^{M}_s=\omega_{_M}^{1/2}\not= 0$, representing the acoustic wave attributed to 
the density perturbation of these massive pairs of normal and dark matter particles. 

Equations (\ref{rden}) and (\ref{pden}) are not mathematically convergent for $k\rightarrow \infty$, due to the particle energy spectrum (\ref{pairspe}) 
$\epsilon_k\rightarrow \infty$. However, the physical ultraviolet cutoff is the Planck mass $M_{\rm pl }$ and the physical relevant scale is 
the large mass $m$ of particles produced in pair production processes that try to produce as many as possible particles of mass $m$, rather than produce a few particles with large kinetic momentum $k > m$. The reason is that 
the pair production probability $|\beta^{(0)}_k(t)|^2$ is suppressed for large $k$. As will be seen, we consider the productions of very massive 
pairs, namely very large pair mass $m\lesssim M_{\rm pl }$ is close to the Planck scale. Therefore, we introduce 
a physical cutoff $k<k_{\rm cut} = m$, i.e., $z<z_{\rm cut} = \mu$
in Eqs.~(\ref{rden}) and (\ref{pden}). 
 
It is conceivable that the spacetime of the horizon $H$ could produces many 
particles and antiparticles (dark matter and normal matter) 
of different masses $m > H$ and degeneracies $g_d$, 
and their energy densities and pressures contribute to total energy 
density $\rho_{_M}$ (\ref{rden})  pressure $p_{_M}$ (\ref{pden}). 
Henceforth, we simply introduce the unique mass-degeneracy parameter ``$m$'' 
to effectively characterise and describe the total contribution from all kinds 
of particle-antiparticle pairs and their degeneracies to the pairs' number 
density (\ref{nden}), energy density (\ref{rden}) and pressure (\ref{pden}). This parameter is simply called ``effective mass parameter'' and more precise definition will be given later. 
\comment{To end this section, we make some remarks about the pair productions of possible bosonic particle and antiparticles, whose number density 
$n^B_{_M}$ goes to zero for $m/H\gg 1$ and has a spurious divergence 
for $m/H \ll 1$ \cite{lnt2014}. Their quantitative contributions 
to the energy density and pressure of matter content are not 
considered in this article, and postponed  
for future studies.
}

\begin{figure}   
\includegraphics[height=5.5cm,width=7.8cm]{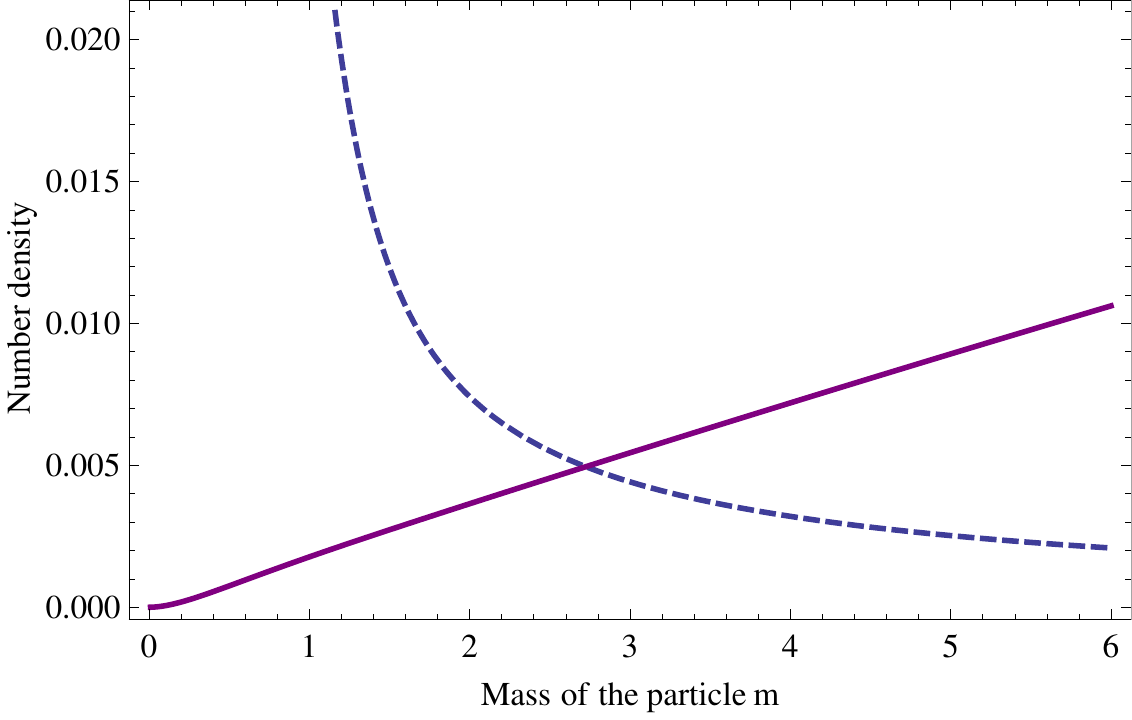}
\vspace{-1em}
\caption{This figure is reproduced from the figure 1 of Ref.~\cite{lnt2014}. 
The violet solid line represents the number density of produced fermion pairs 
$n_{_M}$ (\ref{nden}). 
The blue dashed line  represents the number density of produced boson pairs $n^B_{_M}$ (\ref{ndenB}). They are plotted in terms of the particle mass $m/H$ and $H=1$.}
\label{fbdiff}
\end{figure}

\subsection{Boson-antiboson pair production in De Sitter spacetime}

To end this section, we make some remarks on the 
pair productions of possible bosons $B$ and antibosons $B^\dagger$ 
\begin{eqnarray}
{\mathcal S} \Rightarrow B + \bar B^\dagger,
\label{sproB}
\end{eqnarray}
and contrast it with the pair production of fermions and antifermions. 
The number density $n^B_{_M}$ of produced bosons $B$ and antibosons is given by, see 
for example Refs.~\cite{lnt2014,ekhard},
\begin{eqnarray}
n^B_{_M} &=& \frac{H^3e^{-\pi{\mathcal Im}(\nu)}}{16\pi} I(\mu),\label{ndenB}\\
I(\mu)&\equiv&\int_0^\infty dv \frac{v^2}{\sqrt{v^2+\mu^2}} \Big|vH^{(1)}_{\nu-1}(v)-
\Big[\nu +i\sqrt{v^2+\mu^2}\nonumber\\
&-&\frac{v^2}{2(v^2+\mu^2)}\Big]H^{(1)}_{\nu}(v)\Big|^2 ,~~~~~
\nonumber
\end{eqnarray}
with $\nu\equiv \sqrt{(9/4)-\mu^2}$ and vanishing coupling 
$\xi R B$ term of the Ricci scalar $R$ and scalar field $B$.

Both the 
fermionic pair number density (\ref{nden}) and the bosonic pair 
number density (\ref{ndenB}) are convergent in the ultraviolet regime.
However, their behaviours are quite different, as shown in Fig.~\ref{fbdiff}. 
One finds that the bosonic pair number density $n^B_{_M}$ vanishes for massive pairs $m/H\gg 1$
and has an infrared divergence in the massless pairs $m/H\ll 1$. This implies almost 
no pair productions of subhorizon sized bosonic particles, whose wavelength $m^{-1}$ 
is smaller than the horizon size $H^{-1}$. Only superhorizon sized ($m/H\ll 1$) 
bosonic particles can be significantly produced. 
For this reason, we do not consider in this article 
the pair production (\ref{sproB}) of bosons and antibosons accounting 
for the subhorizon sized matter content $\Omega_{_M}(h)$ and $\omega_{_M}$ 
in the Einstein equations (\ref{fe3}) and (\ref{fcgeqi20}) 
for the Universe horizon evolution. How bosonic pair productions contribute (impact)
to (on) the energy density and pressure of the subhorizon sized 
matter content is postponed for future studies.

On contrary, one finds in Fig.~\ref{fbdiff} that the fermionic pair number density 
$n_{_M}$ vanishes in the massless limit $m/H\ll 1$, namely almost no pair 
productions for superhorizon sized modes. Whereas, the fermionic pair number 
density increases $n_{_M}\propto mH^2$ for massive fermionic pairs $m/H\gg 1$ of 
subhorizon sized fermion modes. 
This means that fermionic pair productions are dominated by well 
subhorizon sized modes, whose wavelength $m^{-1}$ is much smaller than the 
horizon size $H^{-1}$. This is why in this article we actually consider only fermionic 
pair production and particularly massive $m/H\gg 1$ fermion-antifermion ($F\bar F$)
pairs' contributions to the subhorizon sized matter content $\Omega_{_M}(h)$ 
and $\omega_{_M}$ in the Einstein equations (\ref{fe3}) and (\ref{fcgeqi20}).
Moreover, it is worthwhile to mention in advance that for the case of $m\gg H$ 
and $m\gg \tau^{-1}_{_H}$ (\ref{hrate}),  we obtain the 
asymptotical expressions of the fermion pair number density 
$n_{_M} \approx  \chi m H^2$ and energy densities,
\begin{eqnarray}
 \rho_{_M} \approx  2\chi m^2H^2[1+(1/2)(H^2/m^2)],\quad 
\chi\approx 1.85\times 10^{-3},
\label{crucial}
\end{eqnarray}
pressure $p_{_M}\approx (1/6)(H^2/m^2) \rho_{_M}$ and determine the numerical coefficient $\chi$. 
These are crucial expressions, 
which makes the studies of 
the cosmological constant, naturally resultant inflation and cosmic 
coincidence problems to be analytically tractable. 

\red{At the end of this section, we present some discussions on
the non-exponentially suppressed number 
and energy densities (\ref{crucial}) 
of very massive particle-antiparticle pairs production 
in cosmological evolution. They are approximately obtained 
in the static case ``$H=$ const'' and leading order 
of adiabatic approximation in Sec.~\ref{pairs}. 
The reason and demonstration of non-exponentially 
suppressed number and energy densities of superheavy particle production have been given in the pioneering work \cite{CKR1998}. 
There, the authors show that due to nonadiabaticity and discontinuous transition in the cosmic scale factor $a(t)$ evolution, the number and energy densities of superheavy particles produced fall off with a finite power of $H/m$ for $m\gg H$, see Eqs.~(12,14,15) in their article. Besides, the authors show the superheavy particle 
production in cosmologically interesting quantities, 
such as dark matter relic abundance, which is plotted in terms 
of $m/H$, see in their Fig.~2 the solid line for 
the inflationary epoch discontinuously into the matter 
dominate epoch. This situation is very different from the exponentially 
suppressed density ($\sim e^{-m/H}$) of superheavy particles produced 
in static or adiabatically evolutional Universe. 
In our scenario under consideration, the cosmological evolution and pair-production vacuum evolution are certainly 
non-adiabatic and discontinuous, since the cosmic scale 
factor $a(t)$ and its time derivatives are back reacted by 
producing massive pairs and these pairs annihilation and decay. 
It happens particularly in the transitions from one 
epoch to another, which will be discussed below.}
 
\section{\bf Cosmic inflation}\label{inflation}

In this section, we study the cosmic inflation in our scenario
based on (i)  the Universe evolution equations (\ref{fe3}) and (\ref{fcgeqi20});
(ii) the pair productions from the spacetime described by the 
number density 
$n_{_M}$ (\ref{nden}), energy density $\rho_{_M}$ (\ref{rden}) and 
pressure $p_{_M}$ (\ref{pden}); (iii) the unique mass parameter  
$m$ representing an effective mass scale. 
Before specifying initial conditions and 
finding solutions, we give a general discussion. 
In the absence of matter, i.e., no pair productions, 
Eq.~(\ref{fe3}) shows the Universe undergoing the inflation 
$a\sim \exp Ht=\exp (\Lambda/3)^{1/2} t$ for constant $H=(\Lambda/3)^{1/2}$, 
driven by the positive gravitational potential of the cosmological 
$\Lambda$-term, which can be described by a negative energy 
density $T^{00}_{_\Lambda}=-\rho_{_\Lambda}$ 
and $p_{_\Lambda}=-\rho_{_\Lambda}$ in the sense of Eq.~(\ref{emt}).
While, in the presence of the matter, pair productions contribute a positive 
mass-energy density $T^{00}_{_M}=\rho_{_M}$ whose negative gravitational 
potential slows down the inflation. On the other hand, 
pair productions are attributed to
the spacetime horizon $H$ and cosmological $\Lambda$-term. 
We attempt to show how these two dynamics compete and balance 
each other to realize a slowly decreasing $H$ inflation until 
its end, fully satisfying theoretical conditions and agreeing 
with observations.

\subsection{Pre-inflation and inflation}
In our scenario, the classical equations from the 
effective Einstein theory in Secs.~\ref{Ein} and the semi-classical 
framework for the 
pair production in Sec.~\ref{production} 
cannot be applied to the Planck regime of quantum gravity. 
Therefore we discuss cosmic inflation by dividing 
it into two epochs: pre-inflation and inflation with different 
initial conditions (\ref{initH}) and (\ref{initLM}).

In the pre-inflation epoch, we postulate that the initial conditions (\ref{initH}) and (\ref{initLM}) are the characteristic horizon scale $H_{\rm ch}=H_\circ$ 
and scaling factor $a_{\rm ch}=a_\circ$, 
describing a pure spacetime nature without any matter content: 
\begin{eqnarray}
h^2_\circ=\Omega^\circ_{_\Lambda}=1,\quad\Lambda_\circ=3H^2_\circ,\quad 
\Omega^\circ_{_M}=0,
\label{initital0a}
\end{eqnarray}
and the critical density $\rho^\circ_c=3H^2_\circ/(8\pi M^{-2}_{\rm pl})$.
This means that the cosmological term 
$\Omega_{_\Lambda}$ is dominant over the matter $\Omega_{_M}$, the latter is completely negligible. 
Needless to say, the initial value $H_\circ$ is bound to be much smaller than the Planck scale or the reduced Planck scale of the quantum regime, where the effects and details of the quantum gravity and/or Planck transition cannot be ignored.  
Nevertheless, we present a numerical study of the pre-inflation epoch for the initial horizon $H_\circ \lesssim m_{\rm pl}$ 
at the reduced Planck scale, to gain an insight into the pre-inflation 
epoch and its qualitative features. This could be useful information 
for us to seek an effective approach to study this quantum regime at 
the initial horizon being close to the Planck scale $M_{\rm pl}$.

In the inflation epoch,
instead, we assume that the initial conditions (\ref{initH}) and (\ref{initLM}) 
are the characteristic horizon scale $H_{\rm ch}=H_*$ 
and scaling factor $a_{\rm ch}=a_*$, 
and the cosmological term is much larger than the matter content: 
\begin{eqnarray}
h^2_*\gtrsim \Omega^*_{_\Lambda}\gg \Omega^*_{_M},\quad 
h^2_*=1,\quad 
\Lambda_*\lesssim 3H^2_*,
\label{initital0+}
\end{eqnarray}
and the critical density $\rho^\circ_c=3H^2_*/(8\pi M^{-2}_{\rm pl})$.
The initial scale $H_*\ll m_{\rm pl}$ is much smaller than the reduced 
Planck scale $m_{\rm pl}$, and its value is determined in the connection with CMB 
observations.

In terminology, we call the pre-inflation 
epoch ($H_\circ > H > H_*$) 
to distinguish it from the inflation 
epoch ($H_* > H > H_{\rm end}$), 
where the $H_{\rm end}$ represents the inflation ending scale, that 
will be duly discussed and become clear in the next section.

Provided with the initial conditions (\ref{initital0a}) 
or (\ref{initital0+}), using Eqs.~(\ref{nden}), (\ref{rden}) 
and (\ref{pden}), we can numerically or analytically 
calculate the matter content $\Omega_{_M}(h)$ as a function of $h$, 
and solve the Universe evolution equations (\ref{fe3}) and (\ref{fcgeqi20}) 
in the pre-inflation or inflation epoch. In general, 
the nontrivial result $\Omega_{_M}(h)\not=0$ enters the Universe 
evolution equations (\ref{fe3}) and (\ref{fcgeqi20}), dynamically
leading to the decrease of the horizon squared $h^2$ and 
cosmological term $\Omega_{_\Lambda}$. 
In turn $\Omega_{_M}(h)$ changes as a function of $h$. In this way 
we completely determine the variations of 
the horizon $h(x)$, cosmological term 
$\Omega_{_\Lambda}(h)$ and matter content $\Omega_{_M}(h)$ 
in the cosmic inflation.

\subsection{Numerical approach to pre-inflation epoch} 

\begin{figure}   
\includegraphics[height=5.5cm,width=7.8cm]{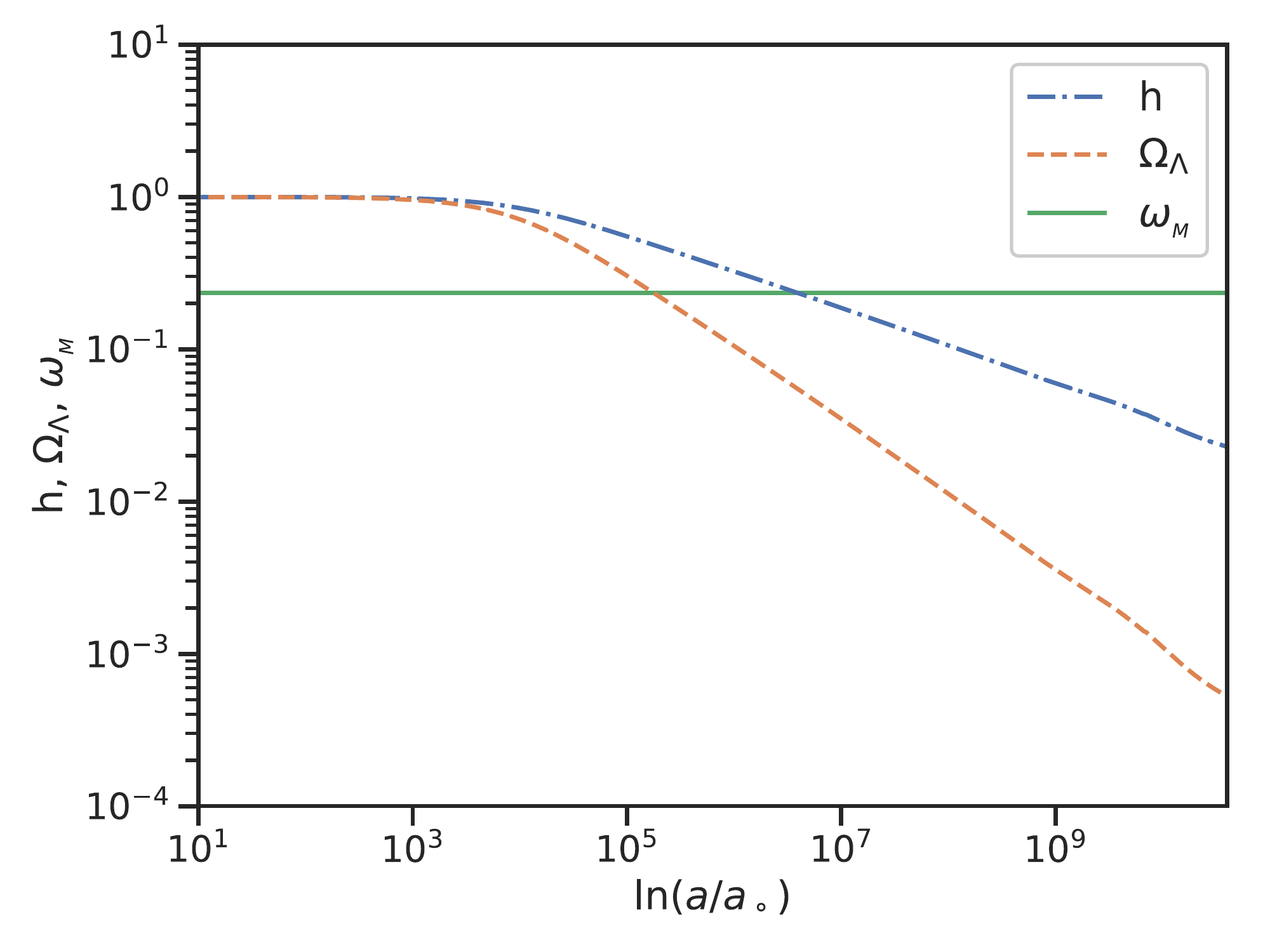}
\vspace{-1em}
\caption{The pre-inflation solution: the inflationary 
characteristics naturally appear, as
$h$ and $\Omega_{_\Lambda}(h)$, $\omega_{_M}=p_{_M}/\rho_{_M}$ 
slowly decrease 
in the $e$-folding number $\ln (a/a_\circ)$. The matter content $\Omega_{_M}(h)$ can be obtained by
$\Omega_{_M}(h)=h^2-\Omega_{_\Lambda}(h)$ of Eq. (\ref{e3}) from 
the plots of $h$ and 
$\Omega_{_\Lambda}(h)$. The characteristic scales and critical density as units 
are given in Eq. (\ref{initital0a}). In this illustration, we adopt the particle 
mass parameter $m=H_\circ=1$.}
\label{finflation}
\end{figure}

For the pre-inflation epoch, selecting the initial horizon scale 
$H_\circ\lesssim m_{\rm pl}$
\comment{
\begin{eqnarray}
H_\circ\lesssim m_{\rm pl},\quad h^2_\circ=1, \quad
h^2_\circ\gtrsim\Omega^\circ_{_\Lambda}\gg\Omega^\circ_{_M},  
\label{initital0a}
\end{eqnarray}
scaling factor $a_\circ$ and the critical density $\rho^\circ_c=3H^2_\circ/(8\pi M^{-2}_{\rm pl})$
}
in the initial conditions (\ref{initital0a}) and the mass parameter $m/H_\circ$, 
we numerically integrate Eqs.~(\ref{fe3},\ref{fcgeqi20}) 
and (\ref{rden},\ref{pden}). As a result, we 
find that the pre-inflation is indeed described by a very slowly decreasing $h^2$ and 
$\Omega_{_\Lambda}(h)$, as illustrated in Fig.~\ref{finflation}. This solution with inflationary characteristics appears naturally without any further {\it ad hoc} adjustment of parameters.  

The physical reasons are clear and follow. The pair 
production (\ref{nden}) is not so rapid that 
the ratio $\Omega_{_M}/\Omega_{_\Lambda}$ 
is very small and slowly increases, therefore $h^2$ and $\Omega_{_\Lambda}$ 
decrease very slowly, see Eq.~(\ref{fcgeqi20}), 
as a function of $e$-folding numbers $\ln (a/a_\circ)$. 
Consequently, in the pre-inflation epoch, we obtain the solution to the cosmological ``constant'', slowly 
varying as an ``area'' law of $H^2$:
\begin{eqnarray}
 \Lambda=3H^2_\circ\Omega_{_\Lambda}(h)\approx 3 H^2\quad {\rm or}\quad 
\Omega_{_\Lambda}(h)\approx h^2.
\label{area1} 
\end{eqnarray}
This result is consistent with Eq.~(\ref{fe3}) 
and negligible $\Omega_{_M}(h)$ in this pre-inflation epoch. 
As shown in our numerical calculations, 
one of them plotted in Fig.~\ref{finflation}, the 
pre-inflation epoch lasts much longer than $\ln (a/a_\circ) > 10^{10}$,
the horizon $h$ and $\Omega_{_\Lambda}(h)$ monotonically decreases, due to the production of 
matter $\Omega_{_M}(h)$, and the ratio  $\Omega_{_M}(h)/\Omega_{_\Lambda}(h)$ 
of the matter term and cosmological term monotonically increases. A 
large amount of matter is expected to be produced by pair 
productions at the scale 
$H_\circ\lesssim m_{\rm pl}$ in this long lasting period. However, 
as already mentioned, such a study of the pre-inflation epoch only 
gives us a qualitative insight into the regimes, whose scales are 
not very much smaller 
than the Planck scale.

Nevertheless, it is allowed to speculate in advance that quantum fluctuation 
modes and acoustic wave (\ref{eosf}) in these regimes could be 
able to exit the horizon and reenter the horizon, later on, imprinting their traces on 
the CMB power spectrum at a larger scale, and/or the power spectrum 
in the nonlinear regimes of forming large scale structure and 
galaxies, even today. 
We would like to point out the sign of the equation of state 
$\omega_{_M}$ (\ref{eosf}) monotonically decreasing 
in Fig.~\ref{finflation}, 
which would relate to observational effects. 
We will come back to this point at the end of the next section.    

\subsection{Analytical approach to inflation epoch}

In the inflation epoch,
we present an analytical and quantitative study of the inflation 
epoch with the characteristic scale and initial horizon 
$H_{\rm ch} = H_*\ll m_{\rm pl}$ being much smaller than the 
reduced Planck scale. We compare our theoretical calculations with
CMB observations.

\subsubsection{Analytical expressions for pair procutions}

Due to the continuous pair productions, 
$H$ and $\Omega_{_\Lambda}$ decrease, 
Equations (\ref{fe3}) and (\ref{fcgeqi20}) for the Universe evolution run into the 
regime of smallness $H/m\ll 1$, 
where it is difficult to perform numerical calculations 
of Hankel functions \cite{num} in Eqs.~(\ref{nden}), (\ref{rden}) and (\ref{pden}) 
for pair productions as functions of $\mu=H/m$. Apart from these numerical difficulties, it is also important to note that in the semi-classical treatment 
of pair productions, the regime $H/m\ll 1$ is physical, 
in the scenes that the wavelengths 
$\lambda^{-1}$ of particles produced are smaller than the radius $H^{-1}$ of the Universe horizon, i.e., $\lambda =m^{-1} \ll H^{-1}$. Therefore there particles 
are well inside the Universe horizon, and their energy-mass content contributes to the Universe evolution. 

We have to find an analytical approach to these  
formulae (\ref{nden}), (\ref{rden}) and (\ref{pden}) 
for calculating pair productions in the regime ($m\gg H$ 
and $m\gg \tau^{-1}_{_H}$). 
We obtain the following asymptotic expressions: 
\begin{eqnarray}
n_{_M} &\approx & \chi m\, H^2,\quad \chi\approx 1.85\times 10^{-3}
\label{aden}\\
\rho_{_M} &\approx & 2\,\chi m^2 H^2(1+ s
),
\label{apden}\\
p_{_M} &\approx&  (s/3) \rho_{_M},
\label{apden0}
\end{eqnarray} 
where $\omega_{_M}=p_{_M}/\rho_{_M} \approx s/3$ and
$s\approx 1/2(H/m)^2\ll 1$. 
We numerically determine $\chi$ value in 
Eqs.~(\ref{aden}) and (\ref{apden}). In addition, 
inserting the damping factor 
$e^{-\sigma(z^2+\mu^2)}$ playing the role of the physical cutoff $z<z_{_{\rm cut}}=\mu$ into Eqs.~(\ref{rden}) and (\ref{pden}), 
we obtain 
Eqs.~(\ref{apden}) and (\ref{apden0}) by the saddle-point approximation of variation with respect to $\sigma$. In the limit of $H/m\ll 1$ and $s\ll 1 $,  
$\rho_{_M} \approx m n_{_M}$ and $p_{_M}\ll 1$, similar to the 
case of massive ``non-reletivistic'' particles. 

We use the pair energy density (\ref{apden}) to effectively define the mass-degeneracy parameter, 
\begin{eqnarray}
\rho^H_{_M} &\approx & 2\,\chi  H^2 m^2;\quad m^2 \equiv 
\sum_fg_d^fm^2_f,
\label{apdenm}
\end{eqnarray}
where $g_d^f$ and $m_f$ are the degeneracy and mass of the 
particle of the flavor $f$, and the $\sum_f$ sums up all flavors produced. The pair-production density (\ref{aden}) 
and rate (\ref{aden0}) show that the pair production process is in favor of massive pairs whose wavelengths are inside the Horizon $H^{-1}$. The inequality $m_f^2>H^2$ implies that the degeneracy $g^f_d$ should be small in the epoch of large $H^2$ value, whereas
it should be large in the epoch of small $H^2$ value. 
Therefore the effective mass-degeneracy parameter $m$, in general, depends on the epoch of the Universe evolution.
The value of this unique mass-degeneracy parameter ``$m$'' 
is determined by observations.

Another important quantity describing the pair-production process 
is the averaged pair-production rate, 
\begin{eqnarray}
\Gamma_M \approx dN/(2\pi dt)\approx (H/2\pi )dN/dx,
\label{prate}
\end{eqnarray}
where $N=n_{_M}H^{-3}/2$ is the number of particles. 
\comment{
\begin{eqnarray}
\Gamma_M \equiv d(n_{_M}V)/dt= d(n_{_M}V)/Hdx,
\label{prate}
\end{eqnarray}
where the volume $V=(4\pi/3) H^{-3}$.}
Using Eq.~(\ref{aden}), we obtain
\begin{eqnarray}
\Gamma_M &\approx& - 
(\chi m/4\pi) (H^{-1}dH/dx) = (\chi m/4\pi) \epsilon
\label{aden0}
\end{eqnarray}
where the $\epsilon$-rate for the Universe evolution 
is defined in Eq.~(\ref{erate0}).

In these analytical formulae, 
the leading order of both $n_{_M}$ (\ref{aden}) and 
$\rho_{_M}$ (\ref{apden}) follows 
the ``area'' law $\propto H^2$, rather than the ``volume'' law $\propto H^3$ 
in Eqs.~(\ref{nden}-\ref{pden}).
The physical picture is that the large number 
(or degeneracies $g_d$) 
$N\sim H^{-1}/m^{-1}\gg  1$ of pairs is produced mainly in the thin 
layer of the width $1/(\chi m)$ on the horizon surface area $H^{-2}$.
This is also in accordance with the spirit of the holographic 
principle \cite{holo}. Otherwise, the number (entropy) $N$ 
of degree freedom would have been vastly 
over-counted for a large horizon size $H^{-1}$, if 
the number density of pairs produced from the spacetime was the 
volume density $n_{_M}\propto H^3$.  
In addition, from Eq.~(\ref{fe3}), we find that the ``area'' 
laws $n_{_M}\propto H^2$ (\ref{aden}) and $\rho_{_M}\propto H^2$ (\ref{apden})  
of pair productions have important physical consequences to the evolution 
of the cosmological $\Lambda$-term, as a function of $H^2$.  
 
We note in advance that in the physical regime ($H/m\ll 1$ and $m\gg \tau^{-1}_{_H}$) 
these analytical expressions (\ref{aden}-\ref{aden0}), 
which approximately describe the Hawking-Parker type process of 
pair-production of particles and antiparticles, are essential for our 
further analyzing each epoch of the Universe evolution:
inflation, reheating, radiation and matter-dominated epochs.    

\subsubsection{Inflation epoch and its end}

To study the inflation epoch, we select the characteristic horizon scale $H_{\rm ch}=H_*$ 
and scaling factor $a_{\rm ch}=a_*$, and the critical density 
$\rho^*_c=3H^2_*/(8\pi M^{-2}_{\rm pl})=3H^2_*m^2_{\rm pl}$, 
moreover
\begin{eqnarray}
h^2_*&=&1 ,\quad  h\equiv H/H_*,\quad H_*\ll m_{\rm pl} ,\nonumber\\
\rho^*_{_M} &=& 2\chi m^2 H^2_*,\quad
\Omega^*_{_M}\equiv \rho^*_{_M}/\rho^*_c=(2/3)\chi (m/m_{\rm pl})^2\label{initital0b}\\
\rho^*_{_\Lambda} &=& \Lambda_*/(8\pi M^{-2}_{\rm pl}),\quad
\Omega^*_{_\Lambda} \equiv \rho^*_{_\Lambda}/\rho^*_c, 
\nonumber\\
\Omega^*_{_\Lambda} &=& h^2_*-\Omega^*_{_M}\gg\Omega^*_{_M},
\label{initital0b+}
\end{eqnarray}
as the initial conditions (\ref{initH}) and (\ref{initLM}) 
for the evolution equations (\ref{fe3}) and (\ref{fcgeqi20}). 
We select {\it a priori}  initial scale  
$H_*\ll m_{\rm pl}$ so that the effects and details of quantum 
gravity and Planck transition could possibly be ignored in the 
inflation epoch, Eqs.~(\ref{fe3}) 
and (\ref{fcgeqi20}) con be approximately valid. We will duly verify the 
condition $H_*\ll m_{\rm pl}$ and Eq.~(\ref{initital0b+}) {\it a posteriori}. 

Using the analytical expressions (\ref{aden}-\ref{aden0}) in the previous section,
the mass-energy content of pairs produced in this epoch is given by
\begin{eqnarray}
\Omega_{_M}\equiv \rho_{_M}/\rho_c^*\approx (2/3)\chi(m/m_{\rm pl})^2(H/H_*)^2(1+s).
\label{omega0}
\end{eqnarray} 
Consequently, Equation (\ref{fcgeqi20}) becomes 
\begin{eqnarray}
dH^2/dx &\approx& - 2\,\chi \,m^2 H^2 (1+\omega_{_M})(1+s),\label{apps}
\end{eqnarray}
yielding the inflationary solution of slowly decreasing $H$
\begin{eqnarray}
H \approx H_*\exp -\chi m^2x=H_*(a/a_*)^{-\chi m^2}.\label{apps0}
\end{eqnarray} 
This is due to the small parameter $\chi m^{2}$ ({\it dimensionless})
\begin{eqnarray} 
\chi m^2\equiv \chi (m/m_{\rm pl})^2\ll 1,\label{dless} 
\end{eqnarray}
that we define here and will use it henceforth, in order to simply notations. 
Readers should not confuse it with the dimensional quantity $m^2$. 
Due to the smallness of the parameter $\chi m^2$, the $H$ is almost constant $H_*$ and
the solution (\ref{apps0}) shows inflationary characteristics.

Because of continuous pair productions, 
the matter $\Omega_{_M}$ makes $H$ 
slowly decreasing (\ref{fcgeqi20}), the inflation slowdown, and 
eventually end at $a=a_{\rm end}$ and $H=H_{\rm end}$. The time 
when the inflation ends can be preliminarily estimated 
by the inflationary rate $H_{\rm end}$ being smaller 
than the averaged pair-production rate $\Gamma_M$ (\ref{aden0}), 
namely
\begin{eqnarray}
H_{\rm end}< \Gamma_M.
\label{infend}
\end{eqnarray}
However, 
this inequality provides the upper bound $H_{\rm end}$ of the horizon $H$ at the 
end of inflation. The value $H_{\rm end}$ should be theoretically 
determined more precisely by studying the dynamical transition from the inflation epoch 
to the reheating epoch, since such transition cannot be instaneous.

We close this section of analytical solution to the inflation by emphasizing the evolution of 
the cosmological ``constant''. In the inflation epoch $H_*>H>H_{\rm end}$, analogously to Eq.~(\ref{area1}) in the pre-inflation epoch, 
the solution to the cosmological ``constant'' is given by the ``area'' law:
\begin{eqnarray}
\Lambda= 3H_*^2\Omega_{_\Lambda} \propto H^2,\label{area2}
\end{eqnarray}
obtained from the fact that 
$\Omega_{_\Lambda}= (H/H_*)^2-\Omega_{_M}$ dominates over 
$\Omega_{_M}\approx (\chi m^2/3)(H/H_*)^2$, i.e., the matter contribution is 
negligible compared with cosmological ``constant'' contributions to the 
inflation of Universe. As will be shown in the next section, the cosmological 
constant term $\Omega_{_\Lambda}$ domination continues up to the inflation end 
defined by $a=a_{\rm end}$ and $H=H_{\rm end}$.

\subsection{Comparison with observations}\label{pivot}

Let the characteristic scale and initial scale  $H_{\rm ch}=H_*$ 
of the inflation
correspond to the interested mode of the pivot scale $k_*$ 
crossed the horizon $(c_sk_*=H_*a_*)$ for CMB observations, 
one calculates the scalar, tensor power spectra and their ratio 
\begin{eqnarray}
\Delta^2_{_{\mathcal R}} 
&=& \frac{1}{8\pi^2}\frac{H^2_*}{m^2_{\rm pl}\,\epsilon\,c_s},\quad
\Delta^2_h 
= \frac{2}{\pi^2}\frac{H^2_*}{m^2_{\rm pl}};\nonumber\\ 
r&\equiv& \frac{\Delta^2_h}{\Delta^2_{_{\mathcal R}}}=16\,\epsilon\, c_s,
\label{ps}
\end{eqnarray} 
where 
the quantity $c_s< 1$ is due to the Lorentz symmetry broken by the time dependence of the background \cite{book}.
The deviations of the scalar and tensor power spectra 
from the scale invariance are described by
\begin{eqnarray}
\Delta^{(n)}_{{\mathcal R},~h}&\equiv& \frac{d^n  \ln \Delta_{{\mathcal R},~h} (k)}{d (\ln k)^n}\Big|_{k_*}
\approx \frac{d^n  \ln \Delta_{{\mathcal R},~h} (k_*)}{d x^n}, \nonumber
\end{eqnarray}
evaluated at the pivot scale $k_*$ for $n=1,2,3,\cdot\cdot\cdot$.  
Adopting the conventional definitions and notations, 
to the leading orders we have  
\begin{eqnarray}
n_s-1 &\equiv& \Delta^{(1)}_{_{\mathcal R}}
\approx -2\epsilon -\eta-\kappa,~~ \alpha_s \equiv \Delta^{(2)}_{_{\mathcal R}}
\approx n_s^\prime, ~~~
\label{ns}\\
n_t &\equiv& \Delta^{(1)}_h =-2\epsilon, ~~\,\,\,\,\,\,\,\,\,\,\,\,\,\,\,~~~~
\tilde n_t \equiv \Delta^{(2)}_h \approx n_t^{\prime},~~~ \label{nt}
\end{eqnarray} 
and $\tilde\alpha_s \equiv \Delta^{(3)}_{_{\mathcal R}}
\approx \alpha_s^\prime 
$, where $\eta \equiv \epsilon'/\epsilon$ and  
$\kappa \equiv  c\,'_s/c_s$. The definition of  
derivative is defined as $(\cdot\cdot\cdot)^\prime\equiv d(\cdot\cdot\cdot)/dx$. 

\subsubsection{Determining the characteristic scale of inflation}

In this theoretical framework, using the solution (\ref{apps}) 
and (\ref{apps0}), we can calculate $\epsilon$ (\ref{erate0}) and 
its high-order derivatives in Eqs.~(\ref{ns}) and (\ref{nt}), 
obtaining
\begin{eqnarray}
\epsilon &\equiv& -H'/H |_{k_*}
\approx \chi\, m^2(1+s),\label{ep}\\ 
\eta &\equiv& \epsilon'/\epsilon |_{k_*}
\approx -3\chi\, m^2s\approx -3\, s\,\epsilon,
\nonumber
\end{eqnarray} 
and 
\begin{eqnarray} 
\eta' \approx -3\eta \epsilon^2, ~\epsilon''\approx \eta^2\epsilon-3\eta\epsilon^3, ~\eta'' \approx 9\eta\epsilon^4-6\eta^2\epsilon^2,\nonumber
\end{eqnarray}
which are evaluated at the pivot scale $k_*$. Equation (\ref{ep}) shows 
$\epsilon \ll 1$, then $\eta < \eta^\prime < \eta^{\prime\prime} \ll 1$. 
In the present article of preliminarily studying the spectral indices and their variations, 
we do not discuss the values of $c_s$
and its variation $\kappa$, simply assuming\ that $c_s\lesssim 1$, 
$\kappa\propto\epsilon^\prime $ and $\kappa^\prime\propto \epsilon^{\prime\prime}$ are small for $H_*/m \ll 1$. Therefore, 
from Eqs.~(\ref{ns}) and (\ref{ep}), we obtain
\begin{eqnarray} 
\epsilon \approx \chi m^2;\quad 2\epsilon \approx 1-n_s -\kappa \approx 1-n_s.
\label{fpara}
\end{eqnarray}
In addition, we calculate the high order variations of the spectral 
indexes (\ref{ns}) and (\ref{nt})
\begin{eqnarray}
n'_s< \epsilon ^2\approx (1-n_s)^2/4,\quad
n''_s < \epsilon ^3\approx (1-n_s)^3/8,\nonumber
\end{eqnarray} 
and we need to know the value of the parameter $\kappa = c\,'_s/c_s$ for further parameter constrains. 

\comment{Due to the spontaneously breaking of De Sitter symmetry, Goldstone boson $\pi$ appears and $\kappa' = d \kappa/ dx \approx -4\kappa \epsilon^2$, as functions of $e$-folding numbers $\ln (a/a_\circ)$, where $y^\prime  \equiv d y/dx$ and $y^{\prime\prime}\equiv d^2 y/dx^2$.}
\comment{
and the observational values $n_s\approx 1-2\epsilon \approx 0.96$ and 
$2\epsilon \approx 0.04$ leading to $m =4.63$, i.e., $m=1.13\times 10^{19}
{\rm GeV} \lesssim M_{\rm pl}=1.22 \times 10^{19}$ GeV. If we put $m=M_{\rm pl}$, $\epsilon = 0.047$. Suppose that the inflation ends when the rate (\ref{prate}) is significantly larger than the expansion rate $H$, parameterizing as 
$\Gamma =(3/8\pi)\alpha H$ and $\alpha >(8\pi/3)$,} 

We are in the position of discussing our theoretical results in connection with observations. 
Based on two CMB observational values 
at the pivot scale $k_*=0.\,05\, ({\rm Mpc})^{-1}$ \cite{Planck2018}: 
\begin{enumerate}[(i)]
\item the spectral index
$n_s\approx 0.965$, 
from which we use Eq.~(\ref{ep}) to estimate the unique parameter of mass scale 
\begin{equation}
m =m_*\lesssim 3.08\, m_{\rm pl},
\label{m*}
\end{equation}
by using $2\epsilon\approx 2\chi  m_*^2\lesssim 1-n_s\approx 0.035$ 
for $\epsilon \gg \eta$ and assuming $2\epsilon < \kappa$. 
Here the dimensionless parameter is defined 
as $\chi m_*^2\equiv\chi (m_*/m_{\rm pl})^2$ (\ref{dless}); 
\item
the scalar amplitude $A_s=\Delta^2_{_{\mathcal R}}(k_*)
\approx  2.1\times 10^{-9}$, from which 
we use Eq.~(\ref{ps}) to determine the characteristic scale, i.e., 
the inflation scale
\begin{equation}
H_*=3.15\times 10^{-5}\,(r/0.1)^{1/2}m_{\rm pl}, 
\label{H*}
\end{equation}
and we use Eq.~(\ref{aden0}) to give the pair-production rate 
\begin{equation}
\Gamma^*_M=(\chi m_*/4\pi)\epsilon=7.9\times 10^{-6}m_{\rm pl}, 
\label{Gamma*}
\end{equation}
at the pivot scale $k_*$ for the mode horizon crossing. 
\end{enumerate}
Note that we adopt the CMB observations to fix the value $m_*$ of the unique mass parameter $m$ introduced to represent the effective mass scale of pair productions and contributions to the mass-energy of matter content. As a result, the inflationary 
scale $H_*$ and pair-production rate $\Gamma^*_M$ are also fixed.
The energy-density ratio of pairs and cosmological term energy densities is given by 
\begin{eqnarray}
\frac{\rho^*_{_M}}{\rho^*_{_\Lambda}}
\approx \frac{2\chi (m_*H_*)^2}{3(m_{\rm pl}H_*)^2}
=\frac{2}{3}\chi m^2_*\approx 1.17\times 10^{-2},
\label{ratio*}
\end{eqnarray}
and $H^2_*\approx \rho^*_{_\Lambda}/(3m^2_{\rm pl})$. This  
shows that the pairs' contribution $\rho_{_M}$ to the Hubble horizon (\ref{fe3}) 
is indeed negligible, compared with the cosmological term contribution 
$\rho_{_\Lambda}$ in the inflation epoch. 
Despite its smallness, the pairs' energy density makes the Hubble rate $H$ 
slowly decrease.

\subsubsection{Inflation $e$-folding number and $r-n_s$ relationship }

Inflation is supposed to end when the condition of 
$\Gamma_M > H_{\rm end}$ (\ref{infend}) is satisfied. 
This is a necessary condition, but could not be a 
sufficient condition. 
Nevertheless, we use this necessary condition to give bounds on the 
tensor-to-scalar ratio 
$r$ and the $e$-folding numbers 
$N_{\rm end}=x_{\rm end}=\ln \left(a_{\rm end}/a_*\right)$ 
from the inflationary scale $H_*$ 
corresponding to the pivot scale $k_*$ to the inflation ending scale $H_{\rm end}$. Using the inflationary solution (\ref{apps0}), we obtain
\begin{eqnarray}
H_{\rm end}=H_*\exp -(\epsilon\, N_{\rm end}),
\label{hend}
\end{eqnarray}
in our scenario.
From Eqs.~(\ref{aden0}) 
and (\ref{hend}),  we have for $\Gamma_M > H_{\rm end}$,
\begin{eqnarray}
(\chi m/4\pi)\,\epsilon > H_*\exp -(\epsilon\, N_{\rm end}),\label{end00}
\end{eqnarray} 
where $H_*$ (\ref{H*}), $\chi m^2\approx 2\epsilon$ and 
$2\epsilon \approx 1-n_s$ (\ref{fpara}) for $\eta,\kappa \ll 1$. 
This yields the number $N_{\rm end}$ of $e$-folding before the inflation end 
\begin{eqnarray}
N_{\rm end}=\ln \left(\frac{a_{\rm end}}{a_*}\right)
&>& \frac{2}{1-n_s}\ln\left[\frac{7.91\times 10^{-4}\,(r/0.1)^{1/2}}{ (1-n_s) \, \chi \,(m/m_{\rm pl})}\right]\label{end0}\\
&=& \frac{2}{1-n_s}\ln\left[\frac{1.12\times 10^{-3}\,(r/0.1)^{1/2}}{(1-n_s)^{3/2} \,\chi^{1/2}}\right],
\label{end}
\end{eqnarray}
in the second line the unique mass parameter $m$ 
is replaced by the observed quantity of spectral index $n_s$:  
$(m/m_{\rm pl})=[(1-n_s)/2\chi]^{1/2}$ (\ref{fpara}). As a result, 
being independent of any free parameter, 
Equation (\ref{end}) yields a definite ($n_s-r$)-relationship between the spectral index $n_s$ 
and the scalar-tensor-ratio $r$, 
\begin{eqnarray}
(r/0.1)<7.97\times 10^{5}\chi (1-n_s)^{3} e^{(1-n_s)N_{\rm end}},
\label{endr}
\end{eqnarray}
for a given $N_{\rm end}$ value of 
inflation $e$-folding number.
For the value $n_s=0.965$ and $m=m_*$ (\ref{m*}), 
we find the results $r <0.037,\, 0.052$ for $N_{\rm end}=50,\,60$ 
in agreement with observations \cite{Planck2018}.  
Moreover, to show such an agreement 
we plot the parameter-free ($n_s-r$) relation (\ref{endr}) 
on the Figure 28 of the Planck 2018 results \cite{Planck2018}, showing that
two curves respectively representing $N_{\rm end}=60$ and $N_{\rm end}=50$ are in the blue zone constrained by observational data sets. 

From Eq.~(\ref{hend}), the inflation ending scale $H_{\rm end}$ is given by
\begin{eqnarray}
H_{\rm end}&=&H_*e^{-\chi m_*^2N_{\rm end}}	\nonumber\\
&\approx &H_*e^{-(1-n_s)N_{\rm end}/2} \approx(0.42,0.35)H_*, 
\label{Hend}
\end{eqnarray}
for the $e$-folding number $N_{\rm end}= (50,60)$ and the tensor-to-scalar 
ratio $r=(0.037,0.052)$. The numerical result $N_{\rm end}$ (\ref{end})
depends on the CMB measurements of $r$ and $n_s$.  
Equations (\ref{H*},\ref{ratio*}), and (\ref{Hend}) 
show that the $H$-variation is very small in the inflation epoch, implying  
\begin{eqnarray}
H^2_{\rm end} = \frac{ \rho^{\rm end}_{_\Lambda}  + \rho^{\rm end}_{_M}}{3m^2_{\rm pl}}\gtrsim \frac{ \rho^{\rm end}_{_\Lambda}}{3m^2_{\rm pl}}; \quad 
\frac{\rho^{\rm end}_{_M}}{\rho^{\rm end}_{_\Lambda}}\ll 1.
\label{ratioend}
\end{eqnarray}
Namely, the cosmological term $\rho^{\rm end}_{_\Lambda}\approx 3
m^2_{\rm pl}H^2_{\rm end}$
is still dominant over the 
pair energy density $\rho^{\rm end}_{_M}\approx 2\chi m^2_*H^2_{\rm end}$
 at the inflation end.   

As a result, from Eqs.~(\ref{m*}), (\ref{H*}) and (\ref{hend}), 
{\it a posteriori} we give a consistent check of our assumption 
$H_{\rm end} < H_*\ll m_{\rm pl}$ and $H_*/m_* \lesssim 1.02\times 10^{-5}\ll 1$, and   
have confidence to adopt 
the semi-classical frameworks and equations presented  
in Secs.~\ref{Ein} and \ref{production}. 
It should be mentioned that the criteria $\Gamma_M >H_{\rm end}$ (\ref{infend}) is indicative, we
expect to determine the values $N_{\rm end}$ and $H_{\rm end}$ by 
detailedly studying the dynamical transition from the inflation 
to the reheating epoch in connection with observations, which are however postponed to 
future investigations. 

\begin{figure}   
\includegraphics[height=7.5cm,width=7.8cm]{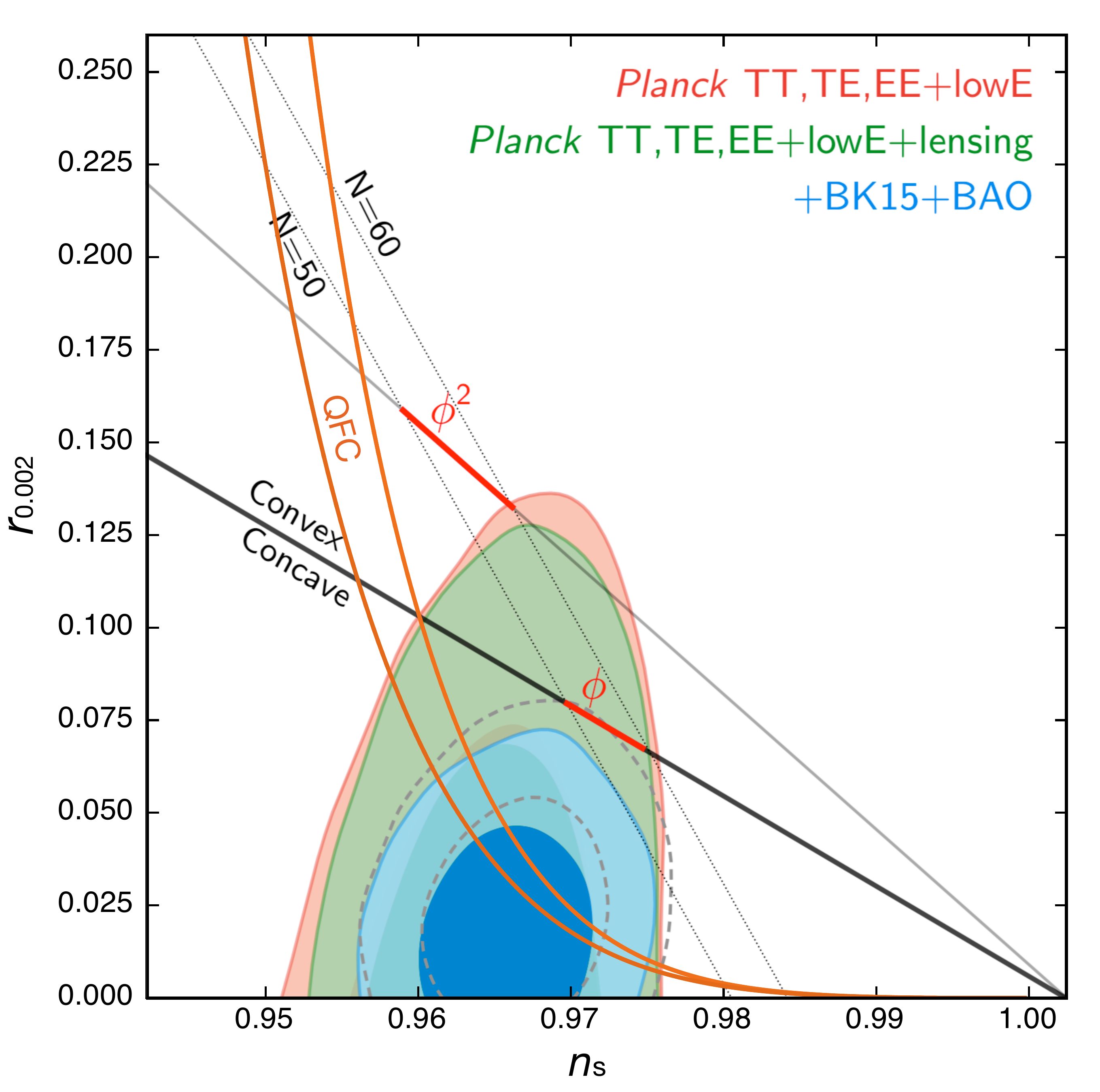}
\vspace{-1em}
\caption{On the Figure 28 of the Planck 2018 results \cite{Planck2018} for constraints on the tensor-to-scalar ratio $r$,  we plot 
the parameter-free ($n_s-r$) relation (\ref{endr}) that shows 
in the observed $n_s$-range, two QFC curves respectively representing $N_{\rm end}=60$ and $N_{\rm end}=50$ are consistently inside the blue zone constrained 
by several observational data sets. The real values of $r$ ratio should be below the curves
due to the nature of inequality (\ref{endr}). As a short notation, the abbreviation 
QFC stands for the model $\tilde\Lambda$CDM studied in this article. 
}
\label{ns-rplot}
\end{figure}

\subsection{Large-scale anomaly and dark-matter acoustic wave}

In order to see any observable physical effect of particle and antiparticle pairs from 
the pre-inflation or inflation epoch,
using Eqs.~(\ref{fe3}) and (\ref{fcgeqi20}), we recast Eqs.~(\ref{ps}) 
and (\ref{erate0}) as 
\begin{eqnarray}
\Delta^2_{_{\mathcal R}} (k) 
&=& \frac{1}{12\pi^2}\frac{H^2{\mathcal R}^{-1}_{_M}}{m^2_{\rm pl}(1+\omega_{_M})c_s}=\frac{1}{8\pi^2}\frac{H^2}{m^2_{\rm pl}\,\epsilon \,c_s},
\label{ps1}\\
\epsilon &=& \frac{3}{2}(1+\omega_{_M}){\mathcal R}_{_M},\label{eps0}
\end{eqnarray}
where the ratio 
\begin{eqnarray}
{\mathcal R}_{_M}	\equiv \frac{\Omega_{_M}}{\Omega_{_\Lambda}+\Omega_{_M}}
= \frac{\rho_{_M}}{\rho_{_\Lambda}+\rho_{_M}}.
\label{rmra}
\end{eqnarray}
Let us examine the evolution  of the scalar spectrum (\ref{ps1}) from 
the ``pre-inflation'' epoch $H_\circ > H\gtrsim H_*$, 
where the equation of state  
$\omega_{_M}\sim 1/3$ (Fig.~\ref{finflation}) for $H_\circ/m \lesssim 1$, 
to the inflation epoch $\omega_{_M}=s/3\ll 1$ 
(\ref{apden0}) for $H_*/m\ll 1$. The evolutions of
$H$ and $\Omega_{_\Lambda,_M}$ 
are very slowly,  
and the ratio ${\mathcal R}_{_M}$ is almost constant, 
${\mathcal R}_{_M}\approx \chi m^2/3$ 
for the inflation epoch $\rho_{_M}\approx \chi m^2H^2$ (\ref{apden}). 
This shows that the scalar spectrum $\Delta^2_{_{\mathcal R}} (k)$ (\ref{ps1}) 
decreases $(1+\omega_{_M})^{-1}\approx 3/4$ at most, due to the  $\omega_{_M}$ variation, as 
the scalar spectrum goes to the large distance scale of the CMB observations, 
exploring high-energy scale of the horizon crossing. 
This probably explains the large-scale anomaly of
the low amplitude of the observed CMB power spectrum at low-$\ell$ multipole, namely the CMB power spectrum 
$\mathcal {D}_l$ drops $\sim 3/4$ at $l=2$. 
These discussions are preliminary and qualitative, 
and further detailed quantitative studies are required.
\comment{Whereas, equation (\ref{eps0}) implies that $\epsilon$ and $r\approx 16\epsilon$ increase,  
$n_s=1-2\epsilon$ decreases, due to $\omega_{_M}$ value increases as the pivot scale $k_*^{-1}$ 
goes to large scales.}  

Moreover, since the equation of state of produced pairs 
is trivial, $\omega_{_M}\not\equiv 0$, and productions 
and annihilations 
of pairs undergo back and forth, 
there could be the acoustic wave of dark-matter and matter density 
perturbations (oscillations) $\delta \rho_{_M}/\rho_{_M}$, 
in the ``pre-inflation'' epoch and the inflation epoch, 
described by the sound velocity 
$c^{M}_s=(\partial p_{_M}/ \partial \rho_{_M})^{1/2}=\omega_{_M}^{1/2}$. 
For the reasons that the most of matter has been produced by pair productions in the pre-inflation and
inflation epochs and the dark matter dominates over the normal matter observed today, we 
suppose that in particle-antiparticle pairs produced in the pre-inflation and inflation epochs, 
there are much more dark-matter particles than normal matter particles. Therefore 
we introduce the abbreviation DAO stands for the dark-matter acoustic oscillations, indicating this 
acoustic wave mainly coming from the dark-matter acoustic oscillations.   
Analogously to metric perturbations, these acoustic waves could exit the horizon and reenter the horizon at large scales. These dark-matter or matter sound waves from the pre-inflation or 
inflation should probably have 
imprinted in the both CMB 
and matter power spectra at large scales of $k_* \sim 10^{-3}{\rm Mpc}^{-1}$. 
This is a phenomenon 
very much similar to baryon acoustic oscillations (BAO). 

We can give here a qualitative description of this DAO phenomenon. 
The acoustic waves come from the DAO in the pre-inflation epoch, whose sound velocity 
$c^{_M}_s=\omega_{_M}^{1/2} < 0.58$ and 
amplitude $\delta\rho_{_M}/\rho_{_M}$ expected to be small, 
because of a small number of relativistic pairs produced. 
They exit the horizon and reenter the horizon again, imprinting their traces 
on the matter power spectra in the large scale structure regime, the nonlinear regime and even today. 
Whereas the acoustic waves come from the DAO in the inflation, whose sound velocity 
$c^{_M}_s=\omega_{_M}^{1/2} \approx (1/6)^{1/2}(m_*/H_*)\sim 10^{-3}-10^{-4}$ and 
amplitude $\delta\rho_{_M}/\rho_{_M}$ expected to be larger than the one from the pre-inflation epoch, 
because of a larger amount of non-relativistic pairs produced. 
They exit the horizon and reenter the horizon again, imprinting their traces on the CMB power spectrum in the last scattering regime of the pivot scale $k_*$. In this article, we do not know the quantitative amplitude $\delta\rho_{_M}/\rho_{_M}$, 
and will present detailed calculations and studies in future publications.  

At the end of this section, we have to mention that 
the results of slowly varying $H$ in the pre-inflation 
epoch (see Fig.~\ref{finflation}) 
and the inflation epoch (\ref{apps0}) in turn justify our approximate calculations 
(\ref{rden}) and (\ref{pden}) by using formulas for a constancy $H$, i.e., 
the adiabatic approximation for the pair-production rate being much 
larger than the rate of the horizon variation. We would like to 
also, emphasize 
that these results of pre-inflation and inflation 
are obtained without any extra field and/or exotic modelling 
beyond the effective Einstein equation and semi-classical 
Hawking-Parker type pair production.

\section{\bf A preliminary discussion of the reheating}\label{entropy}
\comment{To see how to end the inflation, we need to discuss the entropy and particle creation. 
The entropy has two parts: 1. particles and 2. horizon $H$ 
If the reaction rate of $\Lambda$ or space-time 
$H$ creating matter and antimatter which annihilating to $\Lambda$ is faster than Universe expansion, we assume each of them is in thermal equilibrium with the characteristic temperature $T_M$ or $T_H$, where Hawking temperature $T_H=H/2\pi$ of De Sitter space. 
(maybe Universe expansion is faster than some antimatter is out of horizon)}

The inflation epoch ends and reheating epoch starts. The transition and process from one epoch to another epoch cannot be instantaneous and must be very complex. As an example, here we mention one 
microscopic process. In addition to 
the annihilation into the spacetime (\ref{ispro}),  
produced pairs are very massive and decay to relativistic light 
particles. 
In general, the decay rate of massive pairs can be expressed as
\begin{eqnarray}
\Gamma^{\rm decay}_M \propto g^2_{_Y} m
\label{Mdecayr} 
\end{eqnarray}
where $g_{_Y}\sim {\mathcal O}(1)$ is the Yukawa coupling between the massive pairs and
relativistic particles. It is important to note that the decay rate 
$\Gamma^{\rm decay}_M$ (\ref{Mdecayr}) depends not only 
on the Yukawa coupling $g_{_Y}$, but also on the phase space of final particles, 
to which massive pairs decay. 

In the reheating epoch, massive pairs predominately decay to relativistic light particles,
\begin{eqnarray}
\Gamma^{\rm decay}_M >\Gamma_M >H,
\label{Mdecay} 
\end{eqnarray}
and the enormous entropy of a larger amount of relativistic particles is generated. 
The evolution of produced pairs approximately follows the 
conservation of particle number, 
\begin{eqnarray}
\frac{dn_{_M}}{dt}+ 3 Hn_{_M} =  -\Gamma_M^{\rm decay}n_{_M};\quad {\rm i.e.,}\quad  (n_{_M}U^a)_{;\,b} = -\Gamma_M^{\rm decay}n_{_M},
\label{cinsv0d}
\end{eqnarray}
which has to be integrated together with the two basic 
equations (\ref{fe3}) and (\ref{fcgeqi20}).
\comment{
the usual conservation law $d\Omega_{_M}/dx=-3(1+\omega_{_M})\Omega_{_M}$, i.e., is no longer valid, even in the case that particles decouple from the cosmological term. The decay processes can be preliminary described as
\begin{eqnarray}
d\Omega_{_M}/dx \approx -3 \,\omega^{\rm decay}_{_M}\Omega_{_M}, \quad \omega^{\rm decay}_{_M}\equiv \Gamma^{\rm decay}_M/(3H).
\label{decay}
\end{eqnarray}  
} 

Postponing the detailed studies of the reheating epoch to the next article, here
we simply introduce that the reheating end 
is characterized by the time $\tilde t=0$ and scales $a_{\rm ch}=\tilde a$ 
\begin{eqnarray}
H_{\rm ch}=\tilde H,
\quad \tilde\rho_c=3\tilde H^2m^2_{\rm pl},  
\label{initial1}
\end{eqnarray}
and the temperature $\tilde T$. Moreover, we postulate
that at the reheating end the matter term 
$\tilde\Omega_{_M}$ dominates over the cosmological term 
$\tilde \Omega_{_\Lambda}$, namely, 
\begin{eqnarray}
\tilde h^2 \gtrsim\tilde\Omega_{_M}(\tilde H)\gg \tilde\Omega_{_\Lambda}(\tilde H),
\label{initial1+}
\end{eqnarray}
for the possibilities that in the reheating epoch the most amount of cosmological term 
$\Omega_{_\Lambda}$ converts into the matter term $\Omega_{_M}$ that 
is dominant and accounts for the most relevant amount of the matter in the Universe. Some of the massive matter decay to relativistic particles leading to hot Big Bang, others are stable playing the role of cold dark matter. These will be the issues of the next article \cite{xue2019}. 

Equations (\ref{initial1}) and (\ref{initial1+})  are the specific 
initial conditions (\ref{initH}) and (\ref{initLM}) 
for the beginning of the standard cosmology epoch, that we will use 
for integrating basic evolution equations (\ref{fe3}) and (\ref{fcgeqi20}) 
to calculate in next sections the variations of the horizon $H$, 
cosmological term $\Omega_{_\Lambda}$ 
and matter term $\Omega_{_M}$ in the epoch of standard cosmology.

\comment{
As a result, $\Omega_{_\Lambda}$ almost decouples from 
Eq.~(\ref{fcgeqi20}), as if it had been frozen as a ``constant''. Nevertheless $\Omega_{_\Lambda}$ weakly coupling to $\Omega^{\rm l,h}_{_M}$ via $h$ and dominantly governs $h$ again, as decreasing 
$\Omega^{\rm l,h}_{_M}< \Omega_{_\Lambda}$. 
because $\Gamma \ll H$ and $T_M \gg T_H$ in this epoch, particles 
have no enough density and rate to annihilate back to the spacetime.
(iii) $\Gamma < H$, $T_M > T_H$. The spacetime decouples from matter particles, the latter evolves with this huge entropy in standard cosmology, however, there is still interaction between spacetime and matter particles, though they are decoupled each other from thermal equilibrium, and not in equilibrium. The particle matter density value at the heating era is given by Eq.~(\ref{nden}) at $H_{\rm end}$, 
and then follow the law $(a_{\rm end}/a)^4$ to decrease, assume they are a relativistic gas. }

\section{\bf Cosmic coincidence}\label{coincidence}
Our goal in this section is to find 
the variations of the horizon $H$, cosmological term $\Omega_{_\Lambda}$ 
and matter term $\Omega_{_M}$, as well as their relationships 
in the epoch of standard cosmology, by integrating the evolution 
equations (\ref{fe3}) and (\ref{fcgeqi20}) with the initial 
conditions (\ref{initial1}) and (\ref{initial1+}). In order to do this, 
we first need to distinguish two different kinds of matter 
contributions (terms) to the evolution equations (\ref{fe3}) 
and (\ref{fcgeqi20}), because they follow different evolution laws 
and start from the different initial conditions. 

\subsection{Two kinds of matter contributions}

In this theoretical framework, there are two kinds of matter 
contributions to the evolution equations (\ref{fe3}) and (\ref{fcgeqi20}) 
in the standard cosmology epoch.

\subsubsection{$\Lambda$-coupled matter}

The first kind of the matter is called the ``$\Lambda$-coupled'' matter 
and denoted by 
$\Omega^\Lambda_{_M}(h)$ \red{\footnote{\red{The ``$\Lambda$-coupled'' matter is composed of very massive particles, and could be cold dark matter candidates. This will be subject for future studies.}}} and its equation of state 
$\omega^{\Lambda}_{_M}$, indicating its origin 
from the spacetime horizon $H$.
Analogously to massive pairs produced in the inflation,  
the ``$\Lambda$-coupled'' matter is attributed to the 
process of particle and antiparticle ($F\bar F$) pair production 
\red{or annihilation} {\it after} the reheating. 
Their densities, pressure and equation of state 
are computed by Eqs.~(\ref{nden}) and (\ref{rden})-(\ref{eosf}) 
from the initial time $\tilde t=0$ to the final time $t>\tilde t$, 
characterized by another mass parameter $m=\tilde m\not = m_*$ (\ref{m*}).  
This mass parameter $\tilde m$ is unique and introduced to represent 
the effective mass 
scale and degeneracy of pair productions after the reheating epoch.
Its value should be determined by observations.

In the case $H/\tilde m \ll 1$, 
the ``coupled'' matter is approximately represented by the number, energy
densities and equation of state,
\begin{eqnarray} 
n^{\Lambda}_{_M}\approx 2\chi \tilde m H^2,\quad 
\rho^{\Lambda}_{_M}\approx 2\chi \tilde m^2 H^2,\quad {\rm and}\quad 
\omega^{\Lambda}_{_M}\approx 0,
\label{coupled}
\end{eqnarray}
where $H$ denotes the horizon scale of the epoch under consideration. 
These are analogous to Eqs.~(\ref{aden},\ref{apden}) and (\ref{apden0})
in the inflation epoch. However, their initial values are
\begin{eqnarray} 
\tilde n^{\Lambda}_{_M}\approx 2\chi \tilde m \tilde H^2,\quad 
\tilde\rho^{\Lambda}_{_M}\approx 2\chi \tilde m^2 \tilde H^2,\quad {\rm and}\quad 
\omega^{\Lambda}_{_M}\approx 0,
\label{coupled+}
\end{eqnarray}
corresponding to the initial conditions (\ref{initial1}) and (\ref{initial1+}) at the end of the reheating. In the unit of the critical density $\tilde\rho_c$ (\ref{initial1}), we have 
\begin{eqnarray} 
\Omega^{\Lambda}_{_M}(h^2)=\frac{\rho^{\Lambda}_{_M}}{\tilde\rho_c}\approx (2/3)\chi \tilde m^2 h^2,~~{\rm and}~~ 
\tilde \Omega^{\Lambda}_{_M}\approx (2/3)\chi \tilde m^2,
\label{coupled++}
\end{eqnarray}
where $h^2=(H/\tilde H)^2$ and the dimensionless parameter 
$\chi \tilde m^2=\chi (\tilde m/m_{\rm pl})^2\ll 1$, 
similar to Eq.~(\ref{dless}). 
We will check the validity and consistency of this approximation 
$H/\tilde m \ll 1$ in due course. 

\subsubsection{Conventional matter}
The second kind of matter is called the ``conventional'' matter of all 
particles that have been already
produced {\it by} the end of the reheating, referring to the matter content
$\tilde\Omega_{_M}(\tilde H)$ in Eq.~(\ref{initial1+}). This matter content 
is the same as 
the usual matter content studied in the standard cosmology. 
To avoid too many notations, we henceforth use conventional 
notations $\Omega_{_M}$ and $\omega_{_M}=1/3,0$ to represent the 
conventional matter of relativistic or non-relativistic particles, 
unless otherwise stated. These notations are the same as those used for the
massive pairs in the inflation and readers should not be confusing.  

As will be immediately explained below, the conventional matter $\Omega_{_M}$ approximately follows its own conservation
law ($x=\ln a/\tilde a$), 
\begin{eqnarray}
d\Omega_{_M}/dx \approx -3(1+\omega_{_M})\Omega_{_M},\quad \Omega_{_M}(\tilde a)=\tilde\Omega_{_M}\gg \tilde\Omega_{_\Lambda}.
\label{endrh}
\end{eqnarray}
and its evolution is then represented by
\begin{eqnarray}
\Omega_{_M}\approx \tilde\Omega_{_M}\exp -3(1+\omega_{_M})\, x=\tilde\Omega_{_M}
\left(\frac{\tilde a}{ a}\right)^{3(1+\omega_{_M})}.
\label{endrh+}
\end{eqnarray}
These equations are the same as those in the standard cosmology.

\subsection{Coupled equations for $\Omega_{_M}$ and $\Omega_{_\Lambda}$ evolutions }

The total matter content should contain these two kinds of matter contributions,
\begin{eqnarray}
\Omega^{\rm tot}_{_M}=\Omega^{\Lambda}_{_M}(h) + \Omega_{_M}.
\label{totm}
\end{eqnarray}
The basic evolution equations (\ref{fe3}) and (\ref{fcgeqi20}) become
\begin{eqnarray}
h^2 &=& (\Omega_{_M} + \Omega^{\Lambda}_{_M}+ \Omega_{_\Lambda}),
\label{fe3+}\\
\frac{d}{dx}\left(\Omega_{_M} + \Omega^{\Lambda}_{_M}+\Omega_{_\Lambda}\right)
&=&-3(1+\omega_{_M})\Omega_{_M} 
\label{fcg+}\\
&-& 3(1+\omega^\Lambda_{_M}+ \omega^{\rm decay}_{_M})\Omega^\Lambda_{_M},\nonumber
\end{eqnarray}
where the decay ratio $\omega^{\rm decay}_{_M}\equiv\Gamma^{\rm decay}_M/H$ 
is due to the decay (\ref{cinsv0d}) of massive pairs $\Omega^\Lambda_{_M}$ 
into relativistic or non-relativistic particles. 
Equation (\ref{fe3+}) shows that the cosmological term $\Omega_{_\Lambda}$, the
$\Lambda$-coupled matter $\Omega^{\Lambda}_{_M}(h)$ and the conventional matter
$\Omega_{_M}$ are indirectly 
coupled together via the horizon $h^2$. Their variations depend on each 
other via Eq.~(\ref{fcg+}). 

\subsubsection{Indirect interaction between matter and ``dark energy'' via horizon}

Let us consider the epoch of the conventional matter domination: 
\begin{eqnarray}
\Omega_{_M}\gg \Omega_{_\Lambda}\gtrsim 0 ~~{\rm and} ~~
\Omega_{_M}\gg \Omega^{\Lambda}_{_M}\gtrsim 0,\label{case1}
\end{eqnarray}
after the reheating end (\ref{initial1+}) and (\ref{coupled++}). 
At the leading order for the smallness $(\Omega_{_\Lambda}+\Omega^\Lambda_{_M})/\Omega_{_M}\ll 1$, Equations (\ref{fe3+}) and (\ref{fcg+})
becomes $h^2 \approx \Omega_{_M}$ and Eq.~(\ref{endrh}). We then substitute 
the leading-order result $\Omega_{_M}$ into Eqs.~(\ref{fe3+}) and (\ref{fcg+}) to obtain the corrections from $(\Omega_{_\Lambda}+\Omega^\Lambda_{_M})/\Omega_{_M}\ll 1$
for the next leading order,
\begin{eqnarray}
 h^2 &\approx& (\Omega_{_M} + \Omega^{\Lambda}_{_M}+ \Omega_{_\Lambda}),
\label{fe3++}\\
\frac{d}{dx}\left(\Omega^{\Lambda}_{_M}+\Omega_{_\Lambda}\right)
&\approx& -3(1+\omega^\Lambda_{_M}+ \omega^{\rm decay}_{_M})\Omega^\Lambda_{_M},
\label{fcg++}
\end{eqnarray}
where the $\Lambda$-coupled matter $\Omega^\Lambda_{_M}=\Omega^\Lambda_{_M}(h^2)$ 
is calculated by Eq.~(\ref{coupled++}). Equation (\ref{fcg++}) shows that the
decrease \red{or increase} of the cosmological term $\Omega_{_\Lambda}$ 
is due to 
pair production $\Omega^\Lambda_{_M}>0$ \red{or annihilation 
$\Omega^\Lambda_{_M} <0$}. 
If there was no pair production  \red{or annihilation}
$\Omega^\Lambda_{_M}=0$, the cosmological term $\Omega_{_\Lambda}$ 
would be a constant. 

In fact, that Eq.~(\ref{fcg+}) is split into Eqs.~(\ref{endrh}) and (\ref{fcg++}) implies the approximate 
conservation of the conventional matter produced by the end of the 
reheating epoch. The reasons are after the reheating epoch,  
\begin{enumerate}[(i)]
\item the 
pair-production  \red{or pair-annihilation} density $n^\Lambda_{_M}$ (\ref{coupled}) is very small and its contribution to the total matter content (\ref{totm}) 
    is negligible, compared with the conventional matter $\Omega^\Lambda_{_M}$; 
\item the annihilation rate $\Gamma_M$ (\ref{aden0}) of pairs in the conventional matter to the spacetime (\ref{ispro})
is much smaller than the decay rate 
$\Gamma^{\rm decay}_M\propto g^2_{_Y}\tilde m$  
of massive pairs decay (\ref{cinsv0d}) 
to relativistic particles. 
\end{enumerate} 
Thus these two effects (i) and (ii) have negligible 
impacts on the conventional matter $\Omega_{_M}$
and it evolution (\ref{endrh+}). This means that the conventional matter 
$\Omega_{_M}$ does not have direct interactions with the $\Lambda$-term 
$\Omega_{_\Lambda}$ and $\Lambda$-coupled matter $\Omega^\Lambda_{_M}$. 
This can also be seen from Eq.~(\ref{fcg++}), which is independent of the conventional matter $\Omega_{_M}$.

However, the conventional matter $\Omega_{_M}$ indirectly 
couples to the term
$(\Omega^\Lambda_{_M} + \Omega_{_\Lambda})$ through the horizon $h^2$ of 
Eq.~(\ref{fe3++}). Thus, it has impacts on the evolution of the $(\Omega^\Lambda_{_M} + \Omega_{_\Lambda})$ term via the horizon $h$ variation. This can be seen by Eq.~(\ref{fcg++}), whose RHS depends on the horizon $h$, via 
the $\Lambda$-coupled $\Omega^\Lambda_{_M}(h^2)$ (\ref{coupled++}).
As a result, in such a approximation, we obtain the coupled evolution equation 
(\ref{endrh}) or (\ref{endrh+}) and
\begin{eqnarray}
h^2 &\approx& \Omega_{_M} + \Omega_{_\Lambda},
\label{fe3+++}\\
d \Omega_{_\Lambda}/dx
&\approx& -3(1+\omega^\Lambda_{_M}+ \omega^{\rm decay}_{_M})\Omega^\Lambda_{_M}(h^2),
\label{fcg+++}
\end{eqnarray}
where we rewrite $(\Omega^\Lambda_{_M} + \Omega_{_\Lambda})$ as a new
notation $\Omega_{_\Lambda}$ called ``dark energy'', 
since it overall represents the dark energy 
in observations. This means that the dark energy consist of the cosmological term 
and pairs' contribution from the horizon. 
Equations (\ref{fe3+++}) and (\ref{fcg+++}) show 
an indirect interaction of the conventional matter and dark energy through 
the $\Lambda$-coupled matter $\Omega^\Lambda_{_M}(h^2)$ (\ref{coupled++}) 
and the varying horizon scale $h^2$ (\ref{fe3+++}). In addition, 
because of  $\Omega_{_M}\gg \Omega_{_\Lambda}\gtrsim 0$ (\ref{case1}), 
the dark energy $\Omega_{_\Lambda}$ back reaction on the conventional
matter $\Omega_{_M}$ evolution (\ref{endrh+}) is negligible. 

In summary, the conventional matter $\Omega_{_M}$ interacts with 
the dark energy in the following way. 
The conventional matter $\Omega_{_M}$ follows 
evolution (\ref{endrh+}) and impacts on the horizon $h^2$ variation 
(\ref{fe3+++}), which determines the variation 
of the $\Lambda$-coupled matter 
$\Omega^\Lambda_{_M}(h^2)$ (\ref{coupled++}). 
As a result, the dark energy $\Omega_{_\Lambda}$ evolution 
in each epoch after the reheating 
is completely determined by the evolution equation (\ref{fcg+++}) 
and its initial value determined by the transition 
from one epoch to another. 

\subsubsection{Massive pairs decay to relativistic and non-relativistic particles}\label{mpdde}

In order to see how the decay of massive pairs $\Omega^\Lambda_{_M}$ 
impacts on the dark energy in Eq.~(\ref{fcg+++}), we examine  
the so-called decay ratio 
\begin{eqnarray}
\omega^{\rm decay}_{_M}\equiv \Gamma^{\rm decay}_M/H,
\label{decayo}
\end{eqnarray}
which appears in Eqs.~(\ref{fcg+}), (\ref{fcg++}) and (\ref{fcg+++}). 
In units of expanding rate $H$, this ratio effectively describes the 
massive pairs decay to relativistic or non-relativistic particles 
in the radiation or matter dominate epoch. 
The rate $\Gamma^{\rm decay}_{_M}\propto g_{_Y}\tilde m$ (\ref{Mdecay})
of massive pairs decay to particles
depends not only on $g_{_Y}$ and $\tilde m$ but also on the final states
and phase space of particles that they subsequently decay. Therefore the decay rate $\Gamma^{\rm decay}_{_M}$ varies from the radiation dominate epoch to the matter dominated epoch. We are not able to quantitatively 
calculate the ratio $\omega^{\rm decay}_{_M}(h)$ as a function 
of the Hubble horizon $h$ in time. 

Nevertheless, to have an insight into the variation of the ratio $\omega^{\rm decay}_{_M}(h)$ 
(\ref{decayo}) in the transition from one epoch to another, 
we introduce its effective values in the transition respectively.
In the radiation dominated epoch
\begin{eqnarray}
\omega^{\rm decay}_{_M}&\approx& \omega^{\rm decay}_{_{M,R}},
\label{decayor}
\end{eqnarray}
for final decay products being relativistic particles. 
In the matter dominated epoch
\begin{eqnarray}
\omega^{\rm decay}_{_M}&\approx& \omega^{\rm decay}_{_{M,M}},
\label{decayom}
\end{eqnarray}
for final decay products being non-relativistic particles.
The values of $\omega^{\rm decay}_{_{M,R}}$ 
and $\omega^{\rm decay}_{_{M,M}}$ are different. They are expected to be of the 
order of unity and vary smoothly for the following reasons.  
If the expansion rate is much larger than the decay rate, 
$H> \Gamma^{\rm decay}_M$, spacetime generated pairs (\ref{coupled++}) 
have no enough time to 
undergo the decay process, like a ``decoupled'' phenomenon. Instead,
in the ``coupled'' phenomenon $\Gamma^{\rm decay}_M\gtrsim H$ \cite{kolb}, 
the ratio $\omega^{\rm decay}_{_M}\propto  {\mathcal O}(1)$ 
and the pair decay process can relevantly 
couples to the Universe evolution through Eqs.~(\ref{fe3+++})
and (\ref{fcg+++}). 
\comment{
Their difference $\Delta\omega^{\rm decay}_{_M}$ effectively represents the 
$\omega^{\rm decay}_{_M}$-variation  
in the transition from the radiation dominate epoch to the radiation dominate epoch 
\begin{eqnarray}
\Delta\omega^{\rm decay}_{_M}&=& \omega^{\rm decay}_{_{M,M}}- \omega^{\rm decay}_{_{M,R}}>0.\
\label{decayod}
\end{eqnarray}
The property of $\Delta\omega^{\rm decay}_{_M}>0$ is due to the increasing decay rate 
$\Gamma^{\rm decay}_M$ for a larger and recursively generated phase space of final states of particles and their subsequent decays \cite{pdgdecay}, and the increasing horizon size $H^{-1}$.
} 

\subsection{$\Omega_{_\Lambda}-\Omega_{_M}$ relation and cosmic coincidence}

\comment{In Eq.~(\ref{ggg}), to calculate energy content 
$\Omega^\Lambda_{_M}(h)$ or $\rho^\Lambda_{_M}(h)$ 
of pairs produced after the reheating, 
we adopt the asymptotic solutions (\ref{aden}-\ref{apden0}) with the mass 
parameter $\tilde m$, assuming $\tilde m \gg H$. 
Note that $\omega^{\Lambda}_{_M}\approx 0$ for $H/\tilde m\ll 1$, 
see Eqs.~(\ref{apden}) and (\ref{apden0}). Replacing the dimensionless parameter 
$\chi m^2\equiv \chi (m/m_{\rm pl})^2$ by 
$\chi\tilde m^2\equiv \chi (\tilde m/m_{\rm pl})^2\ll 1$ in 
the asymptotic expressions (\ref{apden}) and (\ref{omega0}), we have 
\begin{eqnarray}
\Omega^\Lambda_{_M}(h)\equiv \frac{\rho^{\Lambda}_{_M}}{\tilde \rho_c}\approx\frac{2\chi\tilde m^2H^2}{3\tilde H^2}
= 2 \chi \tilde m ^2 h^2/3,
\label{gg20}
\end{eqnarray}
where $\tilde \rho_c\equiv 3\tilde H^2m^2_{\rm pl}$.}
We are in the position to find the solution to the coupled equations 
(\ref{fe3+++}) and (\ref{fcg+++}), starting from the initial conditions (\ref{initial1}) and (\ref{initial1+})
at the end of the reheating.
Inserting $\Omega^\Lambda_{_M}$ (\ref{coupled++}) into Eq.~(\ref{fcg+++}), 
we obtain
\begin{eqnarray}
\frac{d\Omega_{_\Lambda}}{dx} +\delta\Omega_{_\Lambda}=-\delta\,\Omega_{_M},\quad \delta\equiv 2\,\chi\, \tilde m^2\,(1+\omega^{\rm decay}_{_M})\ll 1,
\label{gg1}
\end{eqnarray}
which can written as $d\Omega_{_\Lambda}/dx =-\delta\,h^2$ showing that the 
dark energy couples to the horizon $h$.
From the RHS $\delta\,\Omega_{_M}$ of this equation (\ref{gg1}), we notice that the ``horizon coupling'' 
$\delta$ between the cosmological term and conventional matter 
term $\Omega_{_M}$ is not zero, but very small. 
Such a horizon coupling is induced from the pair production  
\red{or annihilation} (\ref{coupled}) at the horizon. Besides, 
the term $\delta\,\Omega_{_\Lambda}$ shows that
the initial conditions for this differential equation crucially depend on the 
$\Omega_{_\Lambda}$-value transition from one epoch to another, as will be shown below.  

\subsubsection{$\Omega_{_\Lambda}-\Omega_{_M}$ tracking in the radiation dominated
epoch}

In the standard cosmology, the radiation dominated ``dark'' epoch starts 
at the reheating end. We have the approximate solution to Eqs.~(\ref{gg1}) and (\ref{endrh+}) 
($x=\ln a/\tilde a$ and $\omega_{_M}=1/3$)
\begin{eqnarray}
\Omega_{_\Lambda} &=& \frac{\delta_{_R} \tilde\Omega_{_M}}{4-\delta_{_R}}e^{-4x} +e^{-\delta_{_R}x}\,\tilde{\mathcal C}=\frac{\delta_{_R}}{4-\delta_{_R}}\Omega_{_M}\ll \Omega_{_M},
\label{gg2}\\
\delta_{_R} &\equiv & 2\,\chi\, \tilde m^2\,[1+\omega^{\rm decay}_{_{M,R}}]\ll 1,
\label{gg2'}
\end{eqnarray}
for $\omega^{\rm decay}_{_{M,R}}$ (\ref{decayor}) slowly varying, 
compared with $\Omega_{_M}$-variation (\ref{endrh+}).
In agreement with the conditions (\ref{initial1}) and (\ref{initial1+}), here we 
choose the initial condition at the reheating end $a=\tilde a$ \red{and 
$\tilde \Omega_{_\Lambda}+\tilde\Omega_{_M}=1$}:
\begin{eqnarray} 
\tilde {\mathcal C}=0,~  \tilde \Omega_{_\Lambda}=\delta_{_R}\tilde
\Omega_{_M}/(4-\delta_{_R})\ll \Omega_{_M},
\label{contin} 
\end{eqnarray}
\red{$\tilde\Omega_{_\Lambda}=\delta_{_R} /4$ 
and $\tilde\Omega_{_M}=1-\delta_{_R} /4$},
for the reason that the reheating epoch 
end and radiation dominated ``dark'' epoch 
are radiation dominated, the transitions from one to another should be ``continuous'',
namely both epochs have $\omega_{_M}=1/3$  and the same values of
$\omega^{\rm decay}_{_{M,R}}$. 
However, this is just an argumentation, it needs a detailed numerical
study of the
reheating epoch and its transition to the radiation dominated epoch, since 
the coefficient $\tilde {\mathcal C}$ (\ref{contin}) is the integration over ``continuous''
transitions from the reheating epoch to the radiation dominated epoch. 

Solution (\ref{gg2}) shows that in the radiation dominated epoch, 
the cosmological term is much smaller than the conventional matter term  
$\Omega_{_\Lambda}\ll \Omega_{_M}$.  Most importantly, it shows that 
the evolution of the cosmological term $\Omega_{_\Lambda}$ linearly tracks  
down (follows) the evolution (\ref{endrh+}) of the conventional matter $\Omega_{_M}$.
Here we adopt the terminology ``track down''  used in the discussions of 
Ref.~\cite{track}. Such tracking continues until the Universe reaches 
the radiation-matter equilibrium of characteristic scale 
$a_{\rm eq}$ and $H_{\rm eq}$, 
where the cosmological and conventional matter terms  arrive at their values,
\begin{eqnarray}
\Omega^{\rm eq}_{_\Lambda}=
\frac{\delta_{_R}}{4-\delta_{_R}}\,\Omega^{\rm eq}_{_M}\ll 1,\quad \Omega^{\rm eq}_{_M}
=\Omega_{_M}(a_{\rm eq})\lesssim 1,
\label{equi}
\end{eqnarray}
in units of the corresponding density $\rho_c^{\,\rm eq}=3 H^2_{\rm eq}$. 
We estimate the ratio $(a_{\rm eq}/\tilde a) = (\tilde  T/T_{\rm eq})\sim 10^{15}{\rm GeV}/ 10\,{\rm eV}\sim 10^{23}$, showing that the radiation dominated ``dark'' epoch is a rather long epoch. 

Since $\Omega_{_M}\gg\Omega_{_\Lambda}$, the conventional matter $\Omega_{_M}$ 
mainly contribute to the horizon evolution $h^2 \gtrsim \Omega_{_M}$ (\ref{fe3+++}).
As a consequence, Equation (\ref{gg2}) 
shows that the cosmological constant is small and varies as an ``area'' law at the leading order ${\mathcal O}(\delta_{_R})$,
\begin{eqnarray}
\Omega_{_\Lambda}\approx  (\delta_{_R}/4)\, h^2\approx (\delta_{_R}/4)\, \Omega_{_M}, 
\label{area3}
\end{eqnarray} 
and its contribution to the horizon $h^2$ is negligible 
for $\delta_{_R}\ll 1$. However, it is due to such a small horizon coupling
$\delta_{_R}$ that the cosmological term follows the conventional matter evolution (\ref{endrh+}) from 
the values $(\tilde\Omega_{_\Lambda},\tilde\Omega_{_M})$  
at the reheating end (\ref{contin}) to the values $(\Omega^{\rm eq}_{_\Lambda},\Omega^{\rm eq}_{_M})$
at the radiation-matter equilibrium (\ref{equi}) in this long and dark epoch. 
\comment{and also had been eventually suppressed to the negligible value, even though the value at the reheating end could be large, provided $\chi \tilde m^2\sim 10^{-11}$is small enough for $\tilde m\sim 10^{-4}$, see below. If $\tilde m\sim 10^{-4}$ is this value, the suppression is not very large, what is account is the value $\tilde\Omega_{_\Lambda}\approx 0$ at the reheating end.}

\subsubsection{Cosmological constant in matter-dominated epoch}

We turn to the matter dominated epoch starting from the radiation-matter equilibrium point $a_{\rm eq}$ (\ref{equi}) 
to the present time $a\simeq a_0$ and $(a/a_{\rm eq})\simeq (1+z)\sim 10^4$. This
is a light epoch and it is very short, compared with 
the long dark epoch previously discussed. We have the following 
approximate solution to Eqs.~(\ref{gg1}) and (\ref{endrh+}) ($x=\ln a/a_{\rm eq}$ 
and $\omega_{_M}=0$)
\begin{eqnarray}
\Omega_{_\Lambda} 
&=&\frac{\delta_{_M}}{3-\delta_{_M}}\Omega_{_M} + e^{-\delta_{_M}\,x}{\mathcal C}^{\rm eq}
\label{gg3}\\
\delta_{_M}&=& 2\,\chi\, \tilde m^2\,[1+\omega^{\rm decay}_{_{M,M}}],
\label{gg3'}
\end{eqnarray}
assuming $\omega^{\rm decay}_{_{M,M}}$ (\ref{decayom}) slowly varying, 
compared with $\Omega_{_M}$-variation (\ref{endrh+}).
The coefficient ${\mathcal C}^{\rm eq}$ has to be fixed by matching the 
solution (\ref{gg3}) with the
condition (\ref{equi}) at the radiation-matter equilibrium
\begin{eqnarray}
{\mathcal C}^{\rm eq} &=&
2\,\chi\tilde m^2\,\Delta\omega^{\rm decay}_{_M}\,\Omega^{\rm eq}_{_M}\not=0, 
\label{ceq}\\
\Delta\omega^{\rm decay}_{_M}
&=& \omega^{\rm decay}_{_{M,R}}/4-\omega^{\rm decay}_{_{M,M}}/3-1/12.
\label{ceq+}
\end{eqnarray}
The factor $1/12$ in Eq.~(\ref{ceq+}) is the variation from relativistic particles 
$\omega_{_M}=1/3$ to non-relativistic particles $\omega_{_M}=0$. 
Recalling the decay ratio definition 
$\omega^{\rm decay}_{_M}$ (\ref{decayo}), 
the $\Delta\omega^{\rm decay}_{_M}$ represents the variation of the decay ratio 
$\omega^{\rm decay}_{_M}$ in the transition from the radiation dominated epoch to 
matter dominated epoch. Its value needs a detailed numerical study of integrating 
over all possible phase space
in such a ``discontinuous'' transition. However, we has not yet been able to do 
this calculation, and adopt $\Delta\omega^{\rm decay}_{_M}$ to represent the 
effective variation from $\omega^{\rm decay}_{_{M,R}}$ (\ref{decayor}) to 
$\omega^{\rm decay}_{_{M,M}}$ (\ref{decayom}). 
The property $\Delta\omega^{\rm decay}_{_M}>0$ is due to a larger and recursively generated phase space of final states of particles and their subsequent decays \cite{pdgdecay}. 

In this light epoch,
the solution (\ref{gg3}) shows that the first term decreases as $\Omega_{_M}\approx \,\Omega^{\rm eq}_{_M}(a/a_{\rm eq})^{-3}=
\Omega^{\rm eq}_{_M}(1+z)^{-3}$, and
$\Omega_{_\Lambda}$ fails to track down $\Omega_{_M}$, but approaches to the second term of a slowly varying ``constant''.  
\begin{eqnarray}
\Omega_{_\Lambda} \approx {\mathcal C}^{\rm eq} e^{-\delta_{_M}\,x},
\label{gg3+}
\end{eqnarray}
where $e^{-\delta_{_M}\,x}\approx 1$ very slowly varies 
for $\delta_{_M}\ll 1$. The failure of $\Omega_{_\Lambda}$ tracking $\Omega_{_M}$  
is due to the changes of matter evolution and its particle content. 
This means that until the current epoch, the cosmological term 
$\Omega_{_\Lambda}$ has been almost ``frozen'' to its value (\ref{ceq}) 
depending on the value $\Omega^{\rm eq}_{_M}$ 
at the radiation-matter equilibrium. 
This gives an explanation why the cosmological term is almost constant 
in the current epoch.

\subsubsection{Estimation of \,$\Omega_{_\Lambda}$ and $\Omega_{_M}$ coupling}

In order to estimate the horizon coupling 
$\delta_{_{R,M}}	\approx 2\chi\tilde m^2$, from Eqs.~(\ref{gg3}) and (\ref{ceq}) we obtain the ratio 
\begin{eqnarray}
\Omega_{_\Lambda}/\Omega_{_M}
\approx (\delta_{_M}/3) + 2\,\chi\tilde m^2\,\Delta\omega^{\rm decay}_{_M}\,(1+z)^{3}.
\label{gg5}
\end{eqnarray}
Using current observations $\Omega^0_{_\Lambda}\approx 0.7$ and 
$\Omega^{0}_{_M}\approx 0.3$, correspondingly the redshift $z+1=(a_0/a_{\rm eq})\sim 10^4$ 
at the radiation-matter equilibrium, we obtain 
\begin{eqnarray}
2\chi\tilde m^2 \Delta\omega^{\rm decay}_{_M}&\approx &  (1+z)^{-3}
\Omega_{_\Lambda}/\Omega_{_M}\nonumber\\
&\approx &  (1+10^4)^{-3}
\Omega^0_{_\Lambda}/\Omega^0_{_M}  \approx 2.3\times 
10^{-12},
\label{gg6+}
\end{eqnarray}
where the superscript or subscript ``$0$'' represents the current epoch.
As discussed at the end of Sec.~\ref{mpdde}, $\Delta\omega^{\rm decay}_{_{M,R}}$,
$\Delta\omega^{\rm decay}_{_{M,M}}$ and  $\Delta\omega^{\rm decay}_{_M}$ 
are of the order of unity $\sim {\mathcal O}(1)$, thus 
we estimate the horizon coupling 
$\delta_{_{R,M}}	\approx 2\chi\tilde m^2 \sim {\mathcal O}(10^{-12})$ and 
the mass scale parameter $\tilde m \sim 10^{14}$ GeV 
coinciding with the characteristic scale (temperature) $\tilde T$ 
of the reheating. This value is also consistent with our assumption 
$\tilde m \gg H$ for using the asymptotic solutions (\ref{aden}-\ref{apden0}) 
in these epochs after the reheating.

On the other hand, the results (\ref{equi}), (\ref{ceq}) and (\ref{gg3+})
show that the current value of the cosmological term is the almost 
same as its value at the radiation-matter equilibrium. We use the 
relation  $\rho^0_{_\Lambda}+\rho^0_{_M}=\rho^0_c$ 
of Eq.~(\ref{fe3}) at the present time to estimate 
the current value of dark energy density
\begin{eqnarray}
\rho^0_{_\Lambda}\approx\rho^{\rm eq}_{_\Lambda}= 
\frac{\delta_{_R}}{4}\,\rho^{\rm eq}_{_M}=\frac{\delta_{_R}}{4}\left(\frac{a_0}{a_{\rm eq}}\right)^3\,\rho^0_{_M}\approx \rho^0_c = \frac{3}{8\pi GH_0^{-2}},
\label{xue2000}
\end{eqnarray}
where $\frac{\delta_{_R}}{4}(a_0/a_{\rm eq})^3\sim {\mathcal O}(1)$, 
namely, the current value of the cosmological constant 
$\Lambda_0\propto H_0^2$ \cite{xue2000}.

\subsubsection{Cosmic coincidence of \,$\Omega_{_\Lambda}$ and $\Omega_{_M}$ values}

\begin{figure}   
\includegraphics[height=7.0cm,width=12.0cm]{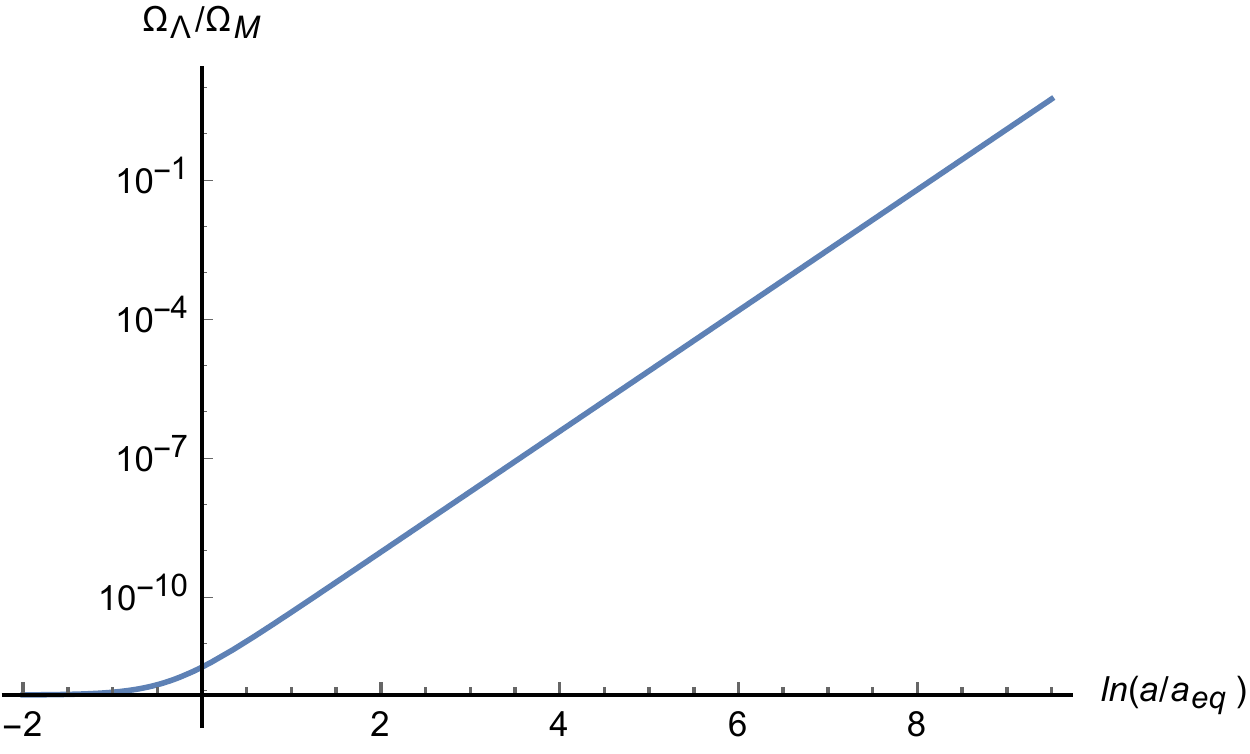}
\vspace{-1em}
\caption{The ratio $\Omega_{_\Lambda}/\Omega_{_M}$ (\ref{gg5+}) is plotted as 
a function of $\ln (a/a_{\rm eq})$, where the scaling factor $a$ 
runs from the reheating end $\tilde a$, through the radiation-matter equilibrium 
$a_{\rm eq}$ to the present time $a_0$, $\tilde a<a_{\rm eq}<a_0$. 
It shows that (i) the tracking-down 
behavior: the ratio is a small constant $\sim 10^{-12}$ for 
$\ln (a/a_{\rm eq})<0$;  (ii) the tracking-down failure occurs around
the radiation-matter equilibrium $\ln (a/a_{\rm eq})=0$; (iii) 
$\Omega_{_\Lambda}\approx {\rm const.}$ (\ref{gg3+}) 
and $\Omega_{_M}\sim (a/a_{\rm eq})^{-3}$, the ratio 
$\Omega_{_\Lambda}/\Omega_{_M}$ increases to 
${\mathcal O}(1)$
at the present time $\ln (a/a_{\rm eq})\approx 9.2$. 
When $\Omega_{_\Lambda}/\Omega_{_M}=1/2$, the Universe turns from 
the deceleration phase to the acceleration phase. The cosmological term 
$\Omega_{_\Lambda}$ will dominate over the matter term $\Omega_{_M}$ in future.}
\label{ccplot}
\end{figure}

To discuss the cosmic coincidence, we use the ratio 
$\Omega_{_\Lambda}/\Omega_{_M}$ which is independent of the 
characteristic scales in different epochs. 
The ratio $\Omega_{_\Lambda}/\Omega_{_M} \approx (\delta_{_R}/4) 
\sim 10^{-12}$  (\ref{area3}) keeps constant, as $\Omega_{_\Lambda}$ 
tracks down $\Omega_{_M}$ from the reheating end $\tilde a$ to 
the radiation-matter equilibrium $a_{\rm eq}\sim 10^{23}\tilde a$. 
This tracking dynamics avoids the fine tuning cosmic  
$\Omega_{_\Lambda}$ and $\Omega_{_M}$ coincidence of the order of 
$(a_{\rm eq}/\tilde a)^4\sim 10^{92}$. Whereas, from the $a_{\rm eq}$ 
to the present time $a_0=(1+z)a_{\rm eq}\sim 10^4 a_{\rm eq}$, we have the 
ratio (\ref{gg3})
\begin{eqnarray}
\Omega_{_\Lambda}/\Omega_{_M}
\approx (\delta_{_M}/3) + \delta_{_M}\,e^{-\delta_{_M}\,x + 3x}.
\label{gg5+}
\end{eqnarray}
where $\Omega_{_M}=\Omega^{\rm eq}_{_M}\, e^{-3x}$ (\ref{endrh+}) 
and $x=\ln (a/a_{\rm eq})$. This ratio 
$\Omega_{_\Lambda}/\Omega_{_M}$ consistently approaches the constant 
${\mathcal O}(10^{-12})$ for $x<0$, i.e., before the radiation-matter 
equilibrium. For an explicit illustration, we plot in Fig.~\ref{ccplot} 
the ratio $\Omega_{_\Lambda}/\Omega_{_M}$ that varies from ${\mathcal O}(10^{-12})$ to ${\mathcal O}(1)$ as a function of the 
scaling factor $\ln (a/a_{\rm eq})$ from the radiation dominated epoch 
to the matter dominated epoch.

These results give us an insight into the issue of the cosmic coincidence 
at present.  
The $\Omega_{_\Lambda}$ and $\Omega_{_M}$ relation shows that 
the cosmic coincidence of $\Omega_{_\Lambda}$ and 
$\Omega_{_M}$ values appear naturally without any extremely fine-tuning, 
since the matter-dominated epoch of $z\sim  10^{3\sim 4}$ 
is much shorter than the radiation dominated epoch 
of $(a^{\rm eq}/\tilde a)\sim 10^{23}$, when 
the $\Omega_{_\Lambda}$ tracks down $\Omega_{_M}$ and the ratio 
$\Omega_{_\Lambda}/\Omega_{_M}$ is constant.  
Otherwise, we would have the cosmic coincidence problem
of incredibly fine-tuning the values $\tilde \Omega_{_\Lambda}$ 
and $\tilde \Omega_{_M}$ 
at the reheating end at the order 
$\sim (10^{-23})^4\times (10^{-4})^3\sim 10^{-104}$, 
to reach their present observational values 
of the same order of magnitude. 

To close this section, we have to mention a few points in the present 
scenario for understanding why the cosmological term is ``constant'' in 
the current epoch, and how the fine-tuning problem of cosmic coincidence 
can be possibly avoided.
\begin{itemize}
\item First, it is necessary to have a small $\Omega_{_\Lambda}$ and $\Omega_{_M}$ 
interaction, whose strength depends on evolution epochs and transitions 
from one to another so that their tracking dynamics proceeds and fails. 
\item Second, despite the detailed numerical analysis necessarily required 
for computing the values $\tilde {\mathcal C}=0$ (\ref{contin}) and 
${\mathcal C}^{\rm eq}\not=0$ (\ref{ceq}), we can be 
sure that the ``continuous'' transition between the reheating epoch 
and the radiation dominated epochs must be different from the 
``discontinuous'' transition between the radiation dominated epoch 
and matter-dominated epoch. This essential difference could be 
the reason why the cosmological ``constant'' $\Omega_{_\Lambda}$ 
evolves (\ref{gg3}) in the matter-dominated epoch very differently 
from its behaviour (\ref{gg2}) of tracking down $\Omega_{_M}$ in the radiation 
dominated epoch. 
Otherwise, $\Delta\omega^{\rm decay}_{_M}=0$ in Eq.~(\ref{ceq+}) and ${\mathcal C}^{\rm eq}=\tilde {\mathcal C}=0$, the 
 cosmological term $\Omega_{_\Lambda}$ would have been tracking down 
 $\Omega_{_M}$-evolution until now, $\Omega_{_\Lambda}\ll \Omega_{_M} \propto h^2$. This is inconsistent with the observations of current Universe acceleration.
\item Third, the estimates of $\delta_{_R}$ (\ref{gg2'}), $\delta_{_M}$ (\ref{gg3'}) and 
${\mathcal C}^{\rm eq}$ (\ref{ceq}) are rather qualitative, since the decay ratio 
$\omega^{\rm decay}$ varies in the ``discontinuous'' transition from the radiation 
dominated epoch to the matter dominated epoch, which takes place in the range 
between $(a_{\rm eq}/a_0)\sim 10^{-4}$ and $(a_{\rm last}/a_0)\sim 10^{-3}$, where 
$a_{\rm last}$ indicates the scaling factor of the last scattering surface. More 
detailed studies are complicate, but 
necessary to reach the quantitative results.
\end{itemize}
Therefore, at this preliminary stage, we would like to treat the 
$\delta_{_M}$ (\ref{gg3'}) and ${\mathcal C}^{\rm eq}$ (\ref{ceq}) 
as parameters to be phenomenologically fixed by observations.

\subsection{Possible connections to observations}

To be in connection with current observations, we rewrite the 
$\Omega_{_\Lambda}$ and $\Omega_{_M}$ relation (\ref{gg3})  
in terms of current observational values $\Omega^0_{_\Lambda}$ and $\Omega^0_{_M}$ 
in units of the critical density $\rho^0_c=3H_0^2$ today,
\begin{eqnarray}
\Omega_{_\Lambda} \approx (\delta_{_M}/3)\,\Omega^0_{_M}(1+z)^3 + \Omega^0_{_\Lambda} (1+z)^{\delta_{_M}},
\label{gg6-}
\end{eqnarray}
and the generalized Friedmann equation (\ref{fe3}) becomes   
\begin{eqnarray}
h^2\approx \,\Omega^0_{_M}(1+z)^{3} 
+ \Omega^0_{_\Lambda} (1+z)^{\delta_{_M}},
\label{gg6--}
\end{eqnarray}
in the matter dominated epoch. \red{The index $\delta_{_M} < 0$ implies that 
the pair annihilation into the spacetime is dominant over the 
pair production from the spacetime. This mostly occurs when $\Omega_{_M} > 
\Omega_{_\Lambda}$. Instead, the index $\delta_{_M} > 0$ implies that 
the pair production from the spacetime is dominant over the pair annihilation into the spacetime. This mostly occurs when $\Omega_{_\Lambda} > \Omega_{_M}$, like the inflation epoch.} 
 
\red{Moreover, 
we have to also take into account the $\Omega_{_M}$ variation due to its coupling with $\Omega_{_\Lambda}$ variation, the possible slow variation of gravitational constant 
and other possible effects. As a result, we recast the generalized Friedmann equation (\ref{e3}) as \cite{xueNPB2015},  
\begin{eqnarray}
h^2= \,\Omega^0_{_M}(1+z)^{3-\delta_{_G}} 
+ \Omega^0_{_\Lambda} (1+z)^{\delta_{_\Lambda}},\label{gg6}
\end{eqnarray}
with two additional parameters $\delta_{_\Lambda}\ll 1$ and $\delta_{_G}\ll 1$.
The parameter $\delta_{_\Lambda}$ relates to $\delta_{_M}$. 
In Eq.~(\ref{gg6}), two parameters $\delta_{_\Lambda}$ 
and $\delta_{_G}$ are constrained, \red{$\delta_{_\Lambda}\delta_{_G} >0$}, 
as required by the generalized conservation law (\ref{cgeqi20}). 
The 
generalized Friedmann equation (\ref{gg6}) shows that the 
characteristic scale for the 
horizon $H$ gets smaller at large $z$ \cite{xueNPB2015}, compared with the $\Lambda$CDM case of $\delta_{_\Lambda}=\delta_{_G}=0$, thus possibly relieves the $H_0$-tension.
Based on observational data, the 
generalized Friedmann equation (\ref{gg6}) has been 
examined and $\delta_{_{\Lambda,G}}$ values have been 
constrained \cite{clement}. It is recently shown \cite{zhangxin} that the generalized Friedmann equation (\ref{gg6}) greatly relieved the $H_0$ tensions 
of the $\Lambda$CDM with some observational 
data \cite{Planck2018,Planck2013,Planck2015,Riess2019,Guo:2018ans}.
}

In particular, how to examine the 
$\Omega_{_\Lambda}$-transition (\ref{gg6-})
from the present ``constant'' $\sim (1+z)^{\delta_{_\Lambda}}$ tracing back 
to the track-down evolution $\sim (1+z)^3$ at the large 
redshift $z\sim 10^{3\sim 4}$. We speculate that such 
$\Omega_{_\Lambda}$-transition 
should induce the peculiar fluctuations of gravitational field that 
imprint on the CMB spectrum, analogously to the integrated 
Sachs-Wolfe effect. 

In addition, Equation (\ref{e2}) gives the turning point $\ddot a=0$ from deceleration $\ddot a <0$ to acceleration $\ddot a >0$, yielding 
$2\Omega_{_\Lambda}=(1+3\omega_{_M})\Omega_{_M}$, i.e.,  $\Omega_{_\Lambda}=\Omega_{_M}/2$ and 
\begin{eqnarray}
(1+z)_{\rm turning} \approx (2\Omega^0_{_\Lambda}/\Omega^0_{_M})^{1/(3-\delta_{_\Lambda}-\delta_{_G})}\approx 1.67,
\end{eqnarray}
and the turning redshift $z_{\rm turning} \approx 0.67$ from the acceleration 
phase to the deceleration phase.

\section{\bf Summary and remarks}

In this article, we emphasize the cosmological $\Lambda$-term 
in the Einstein equation is attributed to the nature of the spacetime rather than the matter. The relevant amount of matter 
is produced from the spacetime horizon, via the process of 
particle and antiparticle pair productions in the pre-inflation, 
inflation and reheating epochs, governed by the cosmological $\Lambda$-term. 
Thereafter, Universe evolution is determined by the time-varying cosmological term, matter term and their interactions via the horizon of the spacetime, 
obeying the Einstein equation and generalized conservation law.

In this theoretical framework, assuming proper initial scales and conditions 
for each epoch of Universe evolution,  
we derive the time evolution of Universe horizon $H$, the cosmological 
term $\Omega_{_\Lambda}(H)$ and matter content $\Omega_{_M}(H)$. 
In the inflation epoch, we calculate the matter content 
$\Omega_{_M}(H)$ that is much smaller than 
$\Omega_{_\Lambda}(H)\propto H^2$. The solution 
naturally leads to the inflation and results agree with observations, 
possibly shows the large scale anomaly of the low amplitude of the CMB power spectrum and the dark-matter acoustic wave. 

We further apply this theoretical framework to study the Universe evolution of the standard cosmology after the reheating. We show the indirect interaction between the cosmological $\Lambda$-term and the matter term through the pair production on the space-time horizon $H$. Such indirect interaction plays the role for the cosmological term $\Omega_{_\Lambda}$ evolution tracking down the matter 
$\Omega_{_M}$ evolution from the reheating end to the radiation-matter equilibrium. 
Afterwards, such a tracking dynamics fails and the cosmological $\Lambda$-term 
varies very slowly up to the present time. This gives a 
possible explanation of why the cosmological term is constant and the possibility of how the problem of cosmic coincidence can be avoided. Besides, due to the matter annihilation to the spacetime, $\Omega_{_\Lambda}$ 
value can increase from the radiation and matter-dominated epochs to the 
$\Omega_{_\Lambda}$-dominated epoch. 

The detailed balance between pair production and annihilation is studied by using the cosmic rate equation \cite{xuereheating}. There, we show that in the reheating
epoch, how the cosmological energy density 
$\rho_{_\Lambda}$ almost completely converts to the matter-energy density 
$\rho_{_M}$, accounting for the most relevant amount of the matter and entropy in the Universe. 
The baryogenesis and magnetogenesis in the reheating epoch are studied in the article \cite{xuehorizon}. There, we show the possibility that the baryogenesis and magnetogenesis are caused by the superhorizon crossing of particle-antiparticle asymmetric perturbations. Besides, we show that these perturbations, as dark-matter acoustic waves, originate in pre-inflation and return to the horizon after the recombination, possibly leaving imprints on the matter power spectrum at large length scales.  

In summary, we provide a possible theoretical scenario to understand the 
issues of the cosmological constant, cosmic inflation, matter origin,
and the cosmic coincidence problem. In this scenario, 
the cosmological term is an attribute of the spacetime horizon, which spontaneously undergoes the pair productions to generate 
the matter term. 
The cosmological and matter terms couple each other via the horizon described by the Einstein equation and generalised conservation law. 
There are other problems to solve in this theoretical framework. 
Further studies are necessarily required and 
a full numerical approach is also inviting.  

\comment{
As have been shown, the Einstein equations for the Friedmann spacetime 
and the process for the matter production consistently describe a 
nonlinear back-reacting process of the Universe evolution.
It is interesting to see whether we can introduce auxiliary scalar fields and potentials 
(interactions) to have an effective action at an appropriate scale to represent 
such complex equations of motions in accordance with observations. This can also be
important for (i) achieving a small-distance (UV) completed and fundamental field theory 
of spacetime and matter; (ii) finding its fixed point, scaling-invariant domain and 
relevant operators to describe observational phenomena at large distances.
}

\comment{    
We emphasize that the area law (\ref{aden}) and (\ref{apden}) are crucial for obtaining the law $\Omega_{_\Lambda}\propto h^2$ \red{up to the radiation-matter equilibrium... }, 
the cosmic inflation and coincidence. 
The initial value 
$\Omega^\circ_{_\Lambda}\propto H^2_\circ$ at the reduced Planck scale should be attributed to the spacetime quantum fluctuation at the Planck scale For some more discussions, see Refs.~\cite{ec_xue2012,xueNPB2015}. \red{this is referee problem.}. The present value 
$\Omega^0_{_\Lambda}\propto H_0^2$ \red{ to be clarified, the present value is the almost same value at decoupling or radiation-matter equilibrium epoch. } is the consequence of $\Omega_{_\Lambda}$ creating and interacting with $\Omega_{_M}$ in the Universe evolution.
} 

To end this article, we make some remarks. Oppositely to the positive mass 
and negative gravitational potential of the matter $\Omega_{_M}$, 
the cosmological term $\Omega_{_\Lambda}$ physically
represents a negative mass-energy (\ref{emt}), 
whose positive potential leads to the 
horizon expansion and pair productions. 
The cosmological term $\Omega_{_\Lambda}$ drives the Universe acceleration as if
an entropy force. 
On the other hand, the pair productions decrease the $\Omega_{_\Lambda}$ and 
``screen'' its positive potential, whereas
these produced pairs increase matter $\Omega_{_M}$ and deepen its negative
potential. As a result, it leads to the 
inflation end and decelerating Universe expansion. 
The positivity of total mass-energy ${\mathcal M}$ of the Universe should be 
expected. 

\section{\bf Acknowledgment}
Author thanks Dr.~Yu Wang for the indispensable numerical assistance of using Python.  

\end{document}